\documentclass[
    prd, twocolumn, superscriptaddress, nofootinbib, amsmath, amssymb,
    aps, floatfix, preprintnumbers
]{revtex4-2}
\usepackage[utf8]{inputenc}
\usepackage[caption=false]{subfig}
\usepackage{enumitem}
\usepackage{setspace}
\usepackage[dvips]{graphicx}
\usepackage[dvipsnames]{xcolor}
\usepackage{colortbl}
\usepackage[
    colorlinks=true,
    linkcolor=blue,
    urlcolor=blue,
    citecolor=blue
]{hyperref}
\usepackage{aas_macros}
\usepackage{ bbold }
\usepackage[capitalise]{cleveref}
\usepackage{siunitx}
\DeclareSIUnit{\year}{yr}

\usepackage{bm}

\newcommand{\beq}{\begin{equation}}
\newcommand{\eeq}{\end{equation}}
\newcommand{\bea}{\begin{eqnarray}}
\newcommand{\eea}{\end{eqnarray}}
\newcommand{\nn}{\nonumber}
\def\beqa{\begin{eqnarray}}
\def\eeqa{\end{eqnarray}}
\newcommand{\no}{\nonumber}

\usepackage{verbatim}

\begin{document}
\title{
Inelastically Decoupling Dark Matter}
\preprint{}

\author{Ronny Frumkin}
\affiliation{Racah Institute of Physics, Hebrew University of Jerusalem, Jerusalem 91904, Israel}
\affiliation{Department of Particle Physics and Astrophysics, Weizmann Institute of Science, Rehovot 7610001,
Israel}

\author{Yonit Hochberg}
\affiliation{Racah Institute of Physics, Hebrew University of Jerusalem, Jerusalem 91904, Israel}
\affiliation{Laboratory for Elementary Particle Physics, Cornell University, Ithaca, NY 14853, USA}

\author{Eric Kuflik}\affiliation{Racah Institute of Physics, Hebrew University of Jerusalem, Jerusalem 91904, Israel}
\affiliation{Laboratory for Elementary Particle Physics, Cornell University, Ithaca, NY 14853, USA}

\author{Binyamin Vilk}
\affiliation{Racah Institute of Physics, Hebrew University of Jerusalem, Jerusalem 91904, Israel}

\date\today

\begin{abstract}\ignorespaces
     We present a new dark matter candidate, the `inELastically DEcoupling Relic' (iELDER), which is a cold thermal relic whose abundance is determined by the freeze out of its inelastic scattering off of bath particles in the presence of $3\to2$
    self-annihilations. The dark matter is predicted to be light, in the ${\cal O}({\rm MeV}-{\rm GeV})$ range,  
     with significant self-annihilations and very weak inelastic couplings to ordinary matter. We demonstrate iELDER dark matter using a $Z_3$-symmetric toy model as well as QCD-like pion theories---the latter showing promising prospects for detection and providing a new benchmark for future searches.  
\end{abstract}

\maketitle

\section{Introduction}\label{sec:intro}
For decades, indirect evidence for the existence of dark matter (DM) has accumulated, however its particle identity remains unknown. Much theoretical and experimental effort has focused on exploring DM as a thermal
relic of the early universe.  The most well-studied thermal relic is the weakly interacting massive particle~(WIMP) scenario, which suggests the DM relic abundance is set by the freeze-out of a $2\rightarrow2$ annihilation process. Experimental exclusion of much of the natural WIMP parameter space has spurred recent works suggesting alternatives to the WIMP (such as Refs.~\cite{Lee:1977ua,Griest:1990kh,Carlson:1992fn,Hall:2009bx,Hochberg:2014dra,Hochberg:2014kqa,DAgnolo:2015ujb,Kuflik:2015isi,Kopp:2016yji,Soni:2016gzf,DAgnolo:2017dbv,Kuflik:2017iqs,DAgnolo:2019zkf,Kim:2019udq,Frumkin:2022ror,Frumkin:2021zng,Frumkin:2025dxq}; for recent reviews, see Refs.~\cite{Battaglieri:2017aum,Asadi:2022njl}). Among these ideas are co-scattering DM~\cite{DAgnolo:2017dbv}, strongly interacting massive particles (SIMPs)~\cite{Hochberg:2014dra,Hochberg:2014kqa} and elastically decoupling relics~(ELDERs)~\cite{Kuflik:2015isi,Kuflik:2017iqs}. 
In the first case, the DM relic abundance is determined by the decoupling of an inelastic DM number-changing scattering process off of bath particles, with a slightly heavier bath particle. In the second case, the $3\rightarrow2$ self-annihilation of DM sets the relic abundance, while an entropy transfer process is open between the dark sector and the bath. In the third case, the DM decouples from the bath while the $3\rightarrow2$ process is still active, and the relic abundance is controlled by the elastic scattering  of the DM off of bath particles. 

\begin{figure}[th!]
    \centering
    \includegraphics[width=0.5\textwidth]{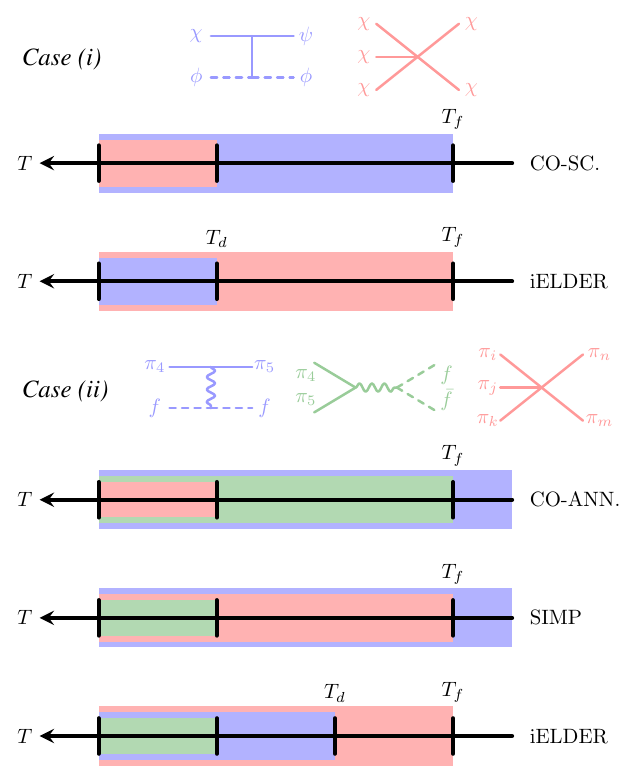}
    \caption{{\bf Schematic setup.} Illustration of the iELDER mechanism and other phases in a general setup (case ({\it i}), {\it top panel}) and a QCD-like pion theory (case ({\it ii}), {\it bottom panel}). Colored areas represent the epochs when the relevant processes are active. $T_d \ (T_f)$ denote temperature of decoupling (freeze out).  {\it Case (i)}: $\psi$ is in chemical equilibrium with the bath due to interactions with the SM (such as $\psi\psi\rightarrow \mathrm{SM}$). Two regimes emerge: iELDER and co-scattering, depending on the order in which the processes decouple. {\it Case (ii):} $\psi$ is one of the dark particles participating in the $3\to2$ process, 
    resulting in  three distinct regimes of co-annihilation, SIMP or iELDER, depending on the order in which the processes decouple. 
    We illustrate this scenario via a QCD-like theory of dark pions. 
    }
    \label{fig:scheme}
\end{figure}

\begin{figure}[ht!]
    \centering
    \subfloat{\includegraphics[width=0.49\textwidth]{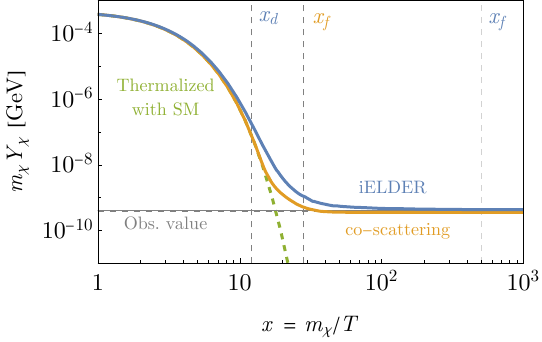} \label{fig:sub1}}
    %
    
   \caption{\textbf{Time evolution.}  
   DM yield times mass 
    $m_\chi Y_\chi$ 
    as a function of $x \equiv m_\chi / T$ with $m_\chi = 10 \ \mathrm{MeV}$, $\Delta=0.1$, $\epsilon = 2\times10^{-8}$ ($2.5\times10^{-8}$) and $\alpha = 5$ $(0.5)$ for iELDER (co-scattering) DM, reproducing the observed abundance.
    We indicate the $x$ values corresponding to iELDER (blue) and co-scattering (orange) decoupling and freeze out. 
    At early times (small $x$), DM remains in thermal equilibrium with the 
    SM bath due to the $\chi\leftrightarrow\psi$ inelastic scattering. In the iELDER regime, this process decouples first, separating the DM bath from the SM bath. The DM then cannibalizes until the $3\rightarrow2$ process freezes out. 
   In contrast, in the co-scattering regime, the $3\rightarrow2$ self-annihilations shuts off first, and DM freeze-out occurs when the inelastic scattering is no longer active. 
   }
    \label{fig:timeevolution}
\end{figure}

In this work we combine main elements from these ideas where the elastic scattering process in SIMPs and ELDERs are replaced with an inelastic scattering process.   
Thus we study a DM candidate  with  two relevant processes for freezeout. The first is  $3\to2$ self-annihilations of DM particles ($\chi_i$),
\beq
\chi_j \chi_k \chi_\ell \to \chi_m \chi_n.
\eeq
The second is an inelastic scattering of DM  into a heavier particle ($\psi$) off of a Standard Model (SM) bath particle~($\phi$), {\it i.e.}
 \beq
\chi_i\phi\leftrightarrow \psi\phi.
\eeq 
The particle $\psi$ has its chemical potential kept to zero during DM freeze-out in one of two ways:

\begin{enumerate}[label=(\roman*),font=\itshape]
\item $\psi$ has number-changing interactions with the SM bath (such as $\psi \psi \rightarrow \rm SM$); or 
\item $\psi$ is one of the $\chi_i$ participating in the $3\to2$ self-annihilations.
\end{enumerate}
In both cases we find a new mechanism of freeze-out where the DM  abundance is controlled by the inelastic scattering process in the presence of $3\to2$ self-annihilations, which we dub an `inELastically DEcoupling Relic', or iELDER. For the latter case  ({\it ii}), we show that this situation generically arises in SIMP pion models, altering the parameter space. Fig.~\ref{fig:scheme} schematically describes the processes involved in setting the relic abundance in both cases and the decoupling hierarchy in the different phases. The top panel matches the first set of assumptions, while the bottom panel relates to the second set. The latter case is illustrated in a QCD-like theory, where we identify the dark sector particles $\chi$ and $\psi$ as dark pions, as will be discussed below.

\begin{figure*}[ht!]
    \centering
    \subfloat{\includegraphics[width=0.48\textwidth]{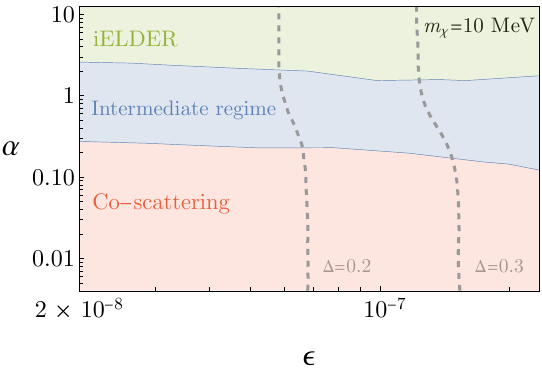} \label{fig:sub1}}
    \subfloat{\includegraphics[width=0.48\textwidth]{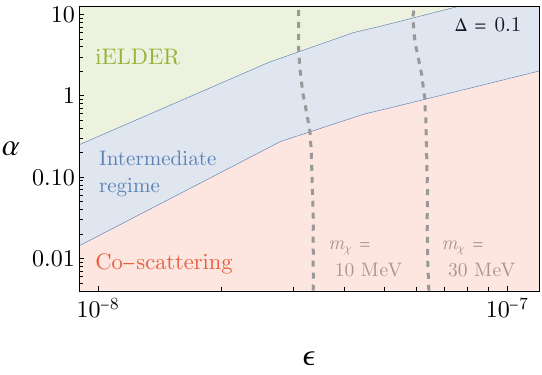} \label{fig:sub2}}
    
    \caption{\textbf{Phases of the setup.} Partition of the $\epsilon-\alpha$ parameter space (defined via Eqs.~\eqref{eq:32CS} and~\eqref{eq:energyT}) to the distinct production mechanisms of co-scattering~(red) and iELDER~(green), with an intermediate regime (blue) marking the smooth transition between the phases.
    {\it Left:} The three phases for a fixed DM mass of $m_\chi=10 \ \mathrm{MeV}$ and varying mass splittings. 
    We show representative curves for $\Delta = 0.2,0.3$. 
    {\it Right:} The same phases for fixed 
    mass splitting $\Delta = 0.1$ and varying DM masses above an MeV. 
    We show representative curves of $m_\chi = 10,30 \ \mathrm{MeV}$. 
    }
    \label{fig:Generalmodel}
\end{figure*}

This paper is organized as follows. Section~\ref{sec:setup} describes a general setup introducing the iELDER DM phase (demonstrated via case ({\it i})), including the relevant Boltzmann equations. 
Section~\ref{sec:z3} presents a toy model that realizes this scenario. Section~\ref{sec:simplest} presents the iELDER phase in  a QCD-like pion model according to the assumptions of case ({\it ii}).  We conclude in Section~\ref{sec:out}.

\section{The iELDER Phase 
}\label{sec:setup}
First we take a general model-independent approach to introduce an iELDER DM phase, under the assumptions of case ({\it i}). Consider a DM candidate $\chi$ with mass $m_\chi$ and a heavier dark particle $\psi$ of mass $m_\psi=m_\chi(1+\Delta)$, where the particles undergo inelastic scattering with a relativistic bath particle $\phi$, $\chi\phi\leftrightarrow \psi\phi$. The $\chi$ particles also undergo self-interactions via a $3\rightarrow2$ annihilation process, $\chi\chi\chi\leftrightarrow\chi\chi$. 

We consider the case where $\psi$ is chemically coupled to the bath at all times through the process $\psi\psi\leftrightarrow\phi\phi$, assuming the process $\chi\chi\leftrightarrow\psi\psi$ can be neglected. The annihilation process, $\chi\psi\leftrightarrow \phi\phi$, always decouples before the number changing inelastic scattering, and so for simplicity we neglect it as well.   At early times, $\chi$ is relativistic and remains in equilibrium with the bath. Later, when the bath temperature drops beneath the mass of $\chi$, $T_\mathrm{bath}<m_\chi$, the $\chi$ number density is exponentially suppressed. If the inelastic scattering  decouples before the $3\to2$ self-annihilations, the DM enters what we denote as the iELDER regime: $\chi$ thermally detaches from the bath and then `cannibalizes'~\cite{Carlson:1992fn}, thus cooling down slower than the bath, and so the DM number density drops slower then if it would have been in thermal equilibrium with the SM bath. 
When the $3\rightarrow2$ process later decouples, the number density of $\chi$ freezes out. Because the number density of $\chi$ only mildly changes throughout the cannibalization epoch, its abundance at freeze out---and hence the relic abundance today---is set by the rate of the inelastic process. 
This is in similar spirit to ELDER DM, though here it is an {\it inelastic} scattering process that is setting the DM abundance. On the other hand, if the $3\to2$ self-annihilations decouple first, DM is in a co-scattering regime \cite{DAgnolo:2017dbv}:  the inelastic scattering is the only number-changing process active, and so when it  decouples it sets the relic abundance. 
Overall, in this setup the DM relic abundance is always set by the inelastic process cross section, but through two different phases,   corresponding to different coupling strengths.

Having described the setup above, we can now write the two coupled Boltzmann equations for the $\chi$ number density $n_\chi$ and energy density $\rho_\chi$,
\beqa\label{eq:BE1}
        \dot{n}_{\chi}+3Hn_{\chi} &=& -\langle\sigma_{3\rightarrow2}v^2\rangle(n_{\chi}^3-n_{\chi}^2n_{\chi}^\mathrm{eq}) \no \\  & & - n_\phi^\mathrm{eq}\langle\sigma_{\chi\rightarrow\psi}v\rangle(n_{\chi} - n_{\chi}^\mathrm{eq})\,,
 \eeqa
and 
\beqa\label{eq:BE2}
\dot{\rho}_{\chi}+3H(\rho_{\chi}+P_{\chi})&=&n_\phi^\mathrm{eq}n_{\psi}\langle \sigma v\cdot E \rangle_{\psi\rightarrow\chi} \no\\ && -  n_\phi^\mathrm{eq}n_{\chi}\langle \sigma v\cdot E \rangle_{\chi\rightarrow\psi}\,.
\eeqa
Here $n_{\phi, \psi}$ are the $\phi$ and $\psi$ number densities respectively, with equilibrium values indicated with superscript `eq'. 
The $3\rightarrow2$ cross section can be computed in a given model in terms of the parameters of the theory; at this stage we use a general parameterization at the mechanism level using dimensional analysis as~\cite{Hochberg:2014dra}
\begin{equation} \label{eq:32CS}
    \langle\sigma v^2 \rangle_{3\rightarrow2} \equiv \frac{\alpha^3}{m_\chi^5}\,,
\end{equation}
with $\alpha$ an effective coupling strength. 

For the inelastic scattering process, the thermally averaged cross section and energy transfer terms are given by
\begin{equation}
    \begin{array}{c}
    n_\phi^\mathrm{eq}n_{\chi} \langle \sigma v \rangle_{\chi\rightarrow\psi} = \\ \int d\Pi_{\chi}d\Pi_{\phi_1}d\Pi_{\psi}d\Pi_{\phi_2}f_{\chi}f_{\phi_1}(2\pi)^4\delta^{(4)}(\Sigma p)\overline{|\mathcal{M}|}_{\chi\rightarrow\psi}^2\
       \end{array}
\end{equation}
and 
\begin{equation}
    \begin{array}{c}
    n_\phi^\mathrm{eq}n_{\chi} \langle \sigma v\cdot E \rangle_{\chi\rightarrow\psi} = \\ \int d\Pi_{\chi}d\Pi_{\phi_1}d\Pi_{\psi}d\Pi_{\phi_2}E_{\chi}f_{\chi}f_{\phi_1}(2\pi)^4\delta^{(4)}(\Sigma p)\overline{|\mathcal{M}|}_{\chi\rightarrow\psi}^2\ \,.
       \end{array}
\end{equation}
Here $\overline{|\mathcal{M}|}_{\chi\rightarrow\psi}^2$ is the averaged  
matrix element squared corresponding to the $\chi\phi\rightarrow\psi\phi$ process, and $f_i$ is the phase space distribution of $i=\chi,\phi$. The backreaction terms in the Boltzmann equations 
can be found 
via detailed balance. 
\begin{figure*}[ht!]
    \centering
    \includegraphics[width=0.48\textwidth]{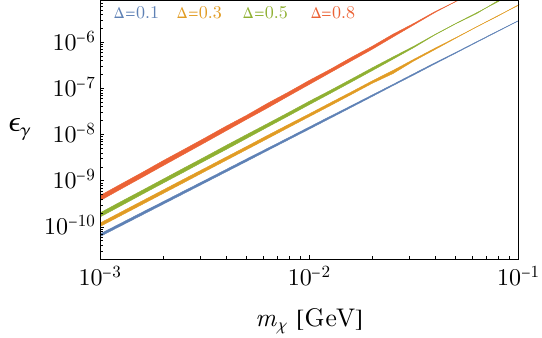}
    \includegraphics[width=0.48\textwidth]{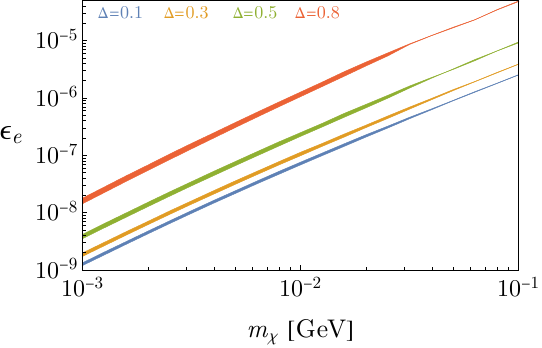}

    \caption{\textbf{$\bm{Z_3}$ model}. The coupling to photons $\epsilon_\gamma$ ({\it left}) and  electrons $\epsilon_e$ ({\it right}) as a function of $m_\chi$ in the $Z_3$ toy model across all DM phases, derived from the numerical solution to the Boltzmann equations, for DM self-couplings  $R\lesssim 4\pi$ 
    and fixed mass splittings $\Delta$.
    }
    \label{fig:Z3modelmassfgamma}
\end{figure*}

The exact calculation of these thermally averaged cross sections and  energy transfers can be performed in specific models.  Assuming a constant matrix element at low DM velocity and taking the leading low temperature and small mass splitting limit, 
we can write a general form for the energy transfer terms,
\begin{eqnarray}\label{eq:energyT}
     \langle \sigma v\cdot E \rangle_{\psi\rightarrow\chi}  \simeq  \frac{\epsilon^2}{m_\chi}\left(1+\frac{\Delta x}{2}\right)\,,\quad
     \epsilon^2\equiv\frac{g_\psi g_\phi \overline{|\mathcal{M}|}^2_{\chi\rightarrow\psi}}{128\pi},
\end{eqnarray}
where $\epsilon$ is a dimensionless parametrization coefficient, $\Delta = (m_\psi-m_\chi)/m_\chi$,  $x=\frac{m_\chi}{T}$  and $g_{\chi, \psi}$ indicating the number of degrees of freedom in each species. Similarly, we can find the thermally averaged cross section, 
\begin{equation}\label{eq:crossS}
     \langle \sigma v \rangle_{\psi\rightarrow\chi}\simeq \frac{\epsilon^2}{m_\chi^2}\left(1+\frac{\Delta x}{2}\right)\\.
\end{equation}
The back-reaction
 forbidden channel is exponentially suppressed in comparison to the non-forbidden channel,
\begin{equation}\label{eq:energyTforbidden}
   \langle \sigma v\cdot E \rangle_{\chi\rightarrow\psi}\propto  e^{-\Delta x}\,, \ \langle \sigma v \rangle_{\chi\rightarrow\psi}\propto  e^{-\Delta x}\, ,
\end{equation}
taking leading order in $\Delta x$.

Eq.~\eqref{eq:crossS} and ~\eqref{eq:energyTforbidden} offer a qualitative understanding of the properties of the energy transfer and cross sections. In the computations below, we include also next order terms. 
Further details are given in Appendix~\ref{app:EnergyTransfer}.

We remark that the use of integrated Boltzmann equations implicitly assumes that $\chi$ particles are in kinetic equilibrium at late times. This assumption can be easily justified if we assume $2\leftrightarrow2$ self scattering with a coupling of order $\alpha$. For reasonably large values, of order $\alpha \gtrsim10^{-3}$, self-scattering will be active during and after freeze out, ensuring kinetic equilibrium throughout the process. 

In what follows, we solve the coupled Boltzmann equations~\eqref{eq:BE1} and~\eqref{eq:BE2} numerically to obtain the time evolution and parameter space of this DM scenario. 
Fig.~\ref{fig:timeevolution} illustrates the time evolution of the DM number density 
in the iELDER and co-scattering regimes 
for DM mass $m_\chi=10$~MeV and mass splitting $\Delta =0.1$
for fixed values of $\epsilon$ and $\alpha$ couplings that produce the observed DM relic abundance. At early times (small $x$), $\chi$ is thermalized with the bath, and so its yield $Y_\chi=n_\chi/s$ follows a Boltzmann distribution and drops exponentially at non-relativistic temperatures. In the iELDER regime~(blue curve), the inelastic scattering process decouples first, at $x_d\sim\mathcal{O}(10)$, $\chi$ detaches from the SM bath and cannibalizes through $3\rightarrow2$ self-annihilations, resulting in a yield that decreases slowly. Later, at $x_f\sim \mathcal{O}(100)$, when the self-annihilations also shut off, the DM abundance freezes out to its observed value. In contrast, in the co-scattering regime~(orange curve), the $3 \to 2$ self-annihilations freeze out first. The DM abundance is then determined by the freezeout of the inelastic co-scattering process, at $x_f\sim {\cal O}(10'{\rm s})$.

Fig.~\ref{fig:Generalmodel} shows the $\epsilon-\alpha$ phase space for a range of DM masses and mass splittings. As predicted,
we find two distinct regimes that reproduce the DM relic abundance,
with a continuous transition between them. We show the partition of the phase space into the different phases---the left panel for fixed DM mass and varying the mass splitting, and the right panel reversed. We also show representative curves for fixed values of $\Delta$~({\it left}) and $m_\chi$~({\it right}), delineating how the couplings transition. Along the upper part of each curve, corresponding to large values of 
$\alpha$, the inelastic scattering process decouples first, creating an iELDER phase. As one flows along the curves to lower values of $\alpha$, the $3 \rightarrow 2$ process decouples first, and the system enters the co-scattering regime. The value of $\epsilon$ only slightly varies between the two phases, because in both cases it is the decoupling of same inelastic scattering process that controls the abundance. 
For iELDER the abundance is determined by the kinetic decoupling of the inelastic scattering, and for co-scattering the abundance is determined by the chemical decoupling of the same process.

\section{Toy $Z_3$ model}\label{sec:z3}

We now present an effective model that realizes the phases presented in Section~\ref{sec:setup}. Consider a DM candidate~$\chi$ that is a complex scalar charged under
an unbroken $Z_3$ symmetry. The effective Lagrangian for such a model is given by 
\begin{equation}
    \mathcal{L} \supset \frac{R}{3!}m_\chi \chi^3+h.c.\ 
\end{equation}
There is, in principle, also a $\lambda_\chi|\chi|^4$ term which will contribute to the full $3\rightarrow2$ process, however for simplicity we set $\lambda_\chi=0$ (see also Ref.~\cite{Kuflik:2017iqs}). This results in a thermally averaged cross section 
\begin{equation}
    \langle\sigma v^2 \rangle_{3\rightarrow2} \simeq 10^{-4} \frac{R^6}{m_\chi^5}\, ,
\end{equation}
that relates $R$ with $\alpha$ defined in Eq.~\eqref{eq:32CS}.
The Lagrangian will  also include a 4-point interaction term, involving the interaction of the DM $\chi$ with a heavier scalar $\psi$ of mass $m_\psi = m_\chi(1+\Delta)$ 
and a pair of photons or fermions. We consider two possibilities,
\begin{equation}\label{eq:photon}
    \mathcal{L}_\gamma \supset \frac{1}{\Lambda^2_\gamma} \chi^\dagger\psi F_{\mu\nu}F^{\mu\nu}+h.c.\,,
\end{equation}
and 
\begin{equation}\label{eq:fermion}
    \mathcal{L}_e \supset \frac{m_e}{\Lambda_e^2} \chi^\dagger \psi \bar{e}e+h.c.\,,
\end{equation}
with $e$ the electron; the generalization to other fermions is straightforward.

This $Z_3$ model yields Boltzmann equations of similar form to Eqs.~\eqref{eq:BE1} and \eqref{eq:BE2}, with $\phi=\gamma$ or $e$. 
If coupling to photons,  the  dimensionless coefficient $\epsilon^2$ related to the rate of the inelastic scattering as defined in Eqs.~\eqref{eq:energyT} and~\eqref{eq:crossS}~is
\begin{equation}
    \epsilon^2_\gamma\simeq \frac{m_\chi^4}{32\pi\Lambda_\gamma^4}\,,
\end{equation}
and if coupling to electrons,
\begin{equation}
\epsilon_e^2\simeq\frac{3 m_e^4}{16\pi\Lambda_e^4}\,.
\end{equation}

One can now numerically solve the Boltzmann equations for the model. 
 The $\alpha-\epsilon_\phi$ coupling phase space (with $\phi$ either photons or electrons) is similar to that obtained for the general case in Fig.~\ref{fig:Generalmodel}. In Fig.~\ref{fig:Z3modelmassfgamma} we show the values of $\epsilon_\gamma$ ({\it left panel}) and $\epsilon_e$ ({\it right panel}) as a function of $m_\chi$, for various fixed mass splittings, that produce the observed DM relic abundance for perturbative self-couplings $R$ across all phases.
The allowed couplings are given by colored regions that appear as a curve with varying thickness. The thickness of the colored region grows for larger mass splittings $\Delta$, as a result of a larger intermediate phase between the co-scattering and iEDLER regimes, namely a larger difference between the solutions in these two phases. For $m_\chi \gtrsim 40$ MeV, the allowed variation in the values of the inelastic couplings $\epsilon_\gamma$ and $\epsilon_e$ is restricted because the iELDER phase cannot be found for perturbative self-couplings $R\lesssim 4\pi$.

We do not discuss existing constraints or future projections of this toy model. 
We move to a more realistic and phenomenologically relevant realization of the iELDER phase in QCD-like theories, next.

\begin{figure*}[ht!]
   \centering
    \includegraphics[width=0.48\textwidth]{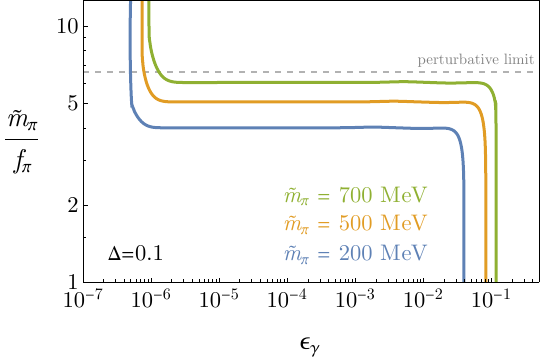} 
    \hfill\includegraphics[width=0.48\textwidth]{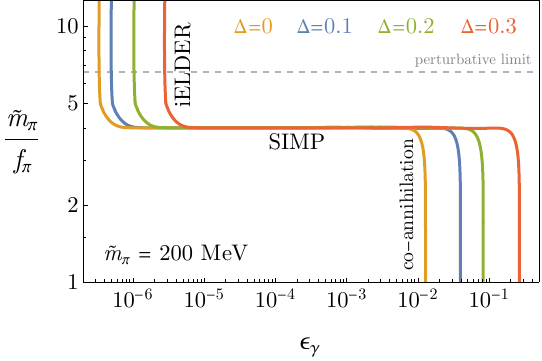}
    
    \caption{\textbf{SIMPlest pion model phase diagram}. $\tilde{m}_\pi/f_\pi$ as the function of $\epsilon_\gamma$  derived from the numerical solution to the Boltzmann Equations for various DM masses $\tilde m_\chi$ and mass splittings $\Delta$, with $m_V = 3\tilde{m}_\pi$ and $\alpha_D=0.1$. Three regimes of the phase space emerge. 
    In all cases, $\pi_4\pi_5$ annihilations decouple before the inelastic scattering process. At small values of $\tilde{m}_\pi/f_\pi$, the $3\to2$ self-annihilations shut off before $\pi_4\pi_5$ annihilations, resulting in a co-annihilation phase. As one flows to larger values of $\tilde{m}_\pi/f_\pi$, $3\to2$ self-annihilations decouple after $\pi_4\pi_5$ annihilations but before the inelastic scattering process shuts off, leading to a SIMP regime. Finally, for even larger values of $\tilde{m}_\pi/f_\pi$, inelastic scattering decouples before $3\to2$ self-annihilations, and an iELDER phase emerges. 
    }
    \label{fig:PionesPlot}
\end{figure*}

\section{QCD-like pion model}\label{sec:simplest}
Having described the iELDER mechanism and demonstrating it in a toy theory,  
we now move to presenting a more realistic model,   
under the assumptions of case~({\it ii}). 
We consider QCD-like theories admitting $3\to2$ self-annihilations, which have been considered in the past as the SIMPlest realization of SIMP DM ~\cite{Hochberg:2014kqa,Hochberg:2015vrg}.
In these theories of dynamical chiral symmetry breaking,  the pseudo-Nambu Goldstone bosons of the theory---which we will collectively call pions---can play the role of DM, with the Wess-Zumino-Witten (WZW) term \cite{Wess:1971yu,Witten:1983tw,Witten:1983tx} generating the $3\rightarrow2$ annihilation process.

\subsection{SIMPlest Setup}
Consider the SIMPlest scenario with an $SU(2)_c$ gauge symmetry and $4$ Weyl fermions $q_i$.
The kinetic terms preserve an $SU(4)$ flavor symmetry, 
\begin{equation}
\mathcal{L}_{\rm kin }= i \bar{q}_i\!\not{\!D} q_i.
\end{equation}
This model leads to chiral symmetry breaking with the order parameter 
\begin{equation}
    \langle q_i \bar{q}_j\rangle=\mu^3J_{ij}\,,
\end{equation}
where $\mu$ is a mass-dimension-one parameter and $J = i\sigma_2 \otimes \mathbb{1} _{2}$
is a $4 \times 4$ anti-symmetric matrix that preserves the flavor subgroup $Sp(4) \subset SU(4)$ \cite{Witten:1983tx,Peskin:1980gc,Preskill:1980mz,Kosower:1984aw}. There are $N_\pi=5$ (Nambu Goldstone boson) pion fields $\pi^a$, $a=1,...5$, 
which will play the role of DM. 
We use the parametrization~\cite{Hochberg:2014kqa}
\begin{equation}
    \Sigma =  \exp(2i\pi/f_\pi)J\,
\end{equation}
for the coset space $SU(4)/Sp(4)$, with 
 \begin{equation}
\pi \;=\;
\begin{pmatrix}
\frac{\pi_{3}}{\sqrt{2}}
  & \frac{\pi_{1} - i\,\pi_{2}}{\sqrt{2}}
  & 0
  & \frac{\pi_{4} - i\,\pi_{5}}{\sqrt{2}}
\\
\frac{\pi_{1} + i\,\pi_{2}}{\sqrt{2}}
  & -\frac{\pi_{3}}{\sqrt{2}}
  & -\frac{\pi_{4} + i\,\pi_{5}}{\sqrt{2}}
  & 0
\\
0
  & -\frac{\pi_{4} + i\,\pi_{5}}{\sqrt{2}}
  & \frac{\pi_{3}}{\sqrt{2}}
  & \frac{\pi_{1} + i\,\pi_{2}}{\sqrt{2}}
\\
\frac{\pi_{4} + i\,\pi_{5}}{\sqrt{2}}
  & 0
  & \frac{\pi_{1} - i\,\pi_{2}}{\sqrt{2}}
  & -\frac{\pi_{3}}{\sqrt{2}}
\end{pmatrix}\,,
 \end{equation}
and $f_\pi$ the pion decay constant. The 5-point interactions, yielding the $3\rightarrow2$ self annihilations, are generated by the WZW term. Under our parameterizations, to leading order in pion fields, we have
\begin{equation} \label{LWZW}
    \mathcal{L}_\mathrm{WZW} = \frac{4}{15\pi^2f_\pi^5}\epsilon^{\mu\nu\rho\sigma}\mathrm{Tr}[\pi\partial_\mu\pi\partial_\nu\pi\partial_\rho\pi\partial_\sigma\pi].
\end{equation}

We take a $U(1)_D$ subgroup of the flavor symmetry to be gauged, corresponding to the generator $Q_D = {\rm diag} \{1,1,-1,-1\}$, which breaks
$SU(4) \to SU(2)\times U(1)_D$. The $U(1)_D$ is spontaneously broken by the VEV of a scalar field $\phi$ that carries charge $-2$. 
This introduces Yukawa couplings that induce symmetry breaking masses,
\begin{eqnarray}
  \mathcal{L}_\phi \supset &&\lambda \phi \epsilon_{ij }Q_i Q_j+\bar{ \lambda}   \phi^* \epsilon_{ij }\bar Q_i \bar Q_j +{\rm h.c.} \nn\\\equiv  &&
 K_{ij}   q_i q_j+{\rm h.c.}
\end{eqnarray}
where $Q=(q_1,q_2), \bar{Q}=(q_3,q_4)$, and 
\begin{equation}
K= \left(\begin{matrix} 0 & m_D & 0 & 0 \\ -m_D& 0 &0 &0 \\ 0 &0&0&\bar{m}_D \\ 0&0&-\bar{m}_D&0\end{matrix} \right)\,,
\end{equation}
with Dirac masses $m_D= \lambda \left<\phi \right>, \ \bar{m}_D= \bar{\lambda}\left<\phi \right>^*  $~\cite{Kulkarni:2022bvh}.  The total quark mass term is  given by
\begin{equation}
\mathcal{L}_{\rm mass }= - \frac{1}{2}M^{ij} q_i q_j+ {\rm h.c.}\,,
\end{equation}
where   
\begin{equation}
    M =  m_QJ - 2K\,
\end{equation}
and $m_Q J$ is the $Sp(4)$-preserving mass contribution. For simplicity, we will consider the case of $m_D=\bar{m}_D$, which assumes an additional charge symmetry $Q\leftrightarrow\bar{Q}$. (For a full study of symmetries of the low energy mesonic spectrum of SIMPlest scenario, see Ref.~\cite{Kulkarni:2022bvh}.)

The pions 
acquire mass,
\begin{eqnarray} \label{firstorder}
    \mathcal{L}_\mathrm{eff} \supset && -\frac{1}{2}\mu^3\mathrm{Tr}M\Sigma  +{\rm h.c.}  \nn\\
   = &&-\frac{m_\pi^2}{4}\mathrm{Tr}\pi^2 + \frac{m_\pi^2}{12f_\pi^2}\mathrm{Tr}\pi^4 + \mathcal{O}\left(\frac{\pi^6}{f_\pi^4}\right)\,,
\end{eqnarray}
where $m_\pi^2\equiv 8 (m_Q\mu^3)/f_\pi^2$ with  the $Sp(4)$-breaking contribution $K$ canceling in the hermitian conjugate.
The symmetry breaking in masses will manifest in the second order term,
\begin{eqnarray} \label{secondorder}
     \Delta\mathcal{L}_\mathrm{eff} &=&  \eta f_\pi^2\mathrm{Tr}M\Sigma M \Sigma +h.c.\nn \\
     &=&-\frac{\eta f_\pi^4m_\pi^4}{4\mu^6}\mathrm{Tr}\pi^2\nn 
     - 64\eta m_D^2\pi_4^2+\frac{1024\eta m_D^2}{3f_\pi^2}\pi_4^2\mathrm{Tr}\pi^2 \nn \\ && +\frac{8\eta f_\pi^2m_\pi^4}{3\mu^6}\mathrm{Tr}\pi^4+\mathcal{O}\left(\frac{\pi^6}{f_\pi^4}\right),
\end{eqnarray}  
where $\eta$ is a dimensionless coefficient. This term introduces a correction to the pion mass, and a 
mass splitting between 
$\pi_4$
and the other pions. The resulting pion masses are
\begin{equation}
    m_{1,2,3,5}^2 = \tilde{m}_\pi^2  \ , \ m_4^2 = \tilde{m}_\pi^2(1+\Delta), \ \Delta = \frac{256\,\eta\, m_D^2}{\tilde{m}_\pi^2},
\end{equation}
where $\tilde{m}_\pi^2 = m_\pi^2\left(1+64\eta\frac{ m_Q^2}{m_\pi^2}\right)$. 
 In addition, terms introduced in Eq.~\eqref{firstorder} and ~\eqref{secondorder}  generate 4-point pion self-scattering. 

The kinetic term also generates terms inducing pion self-scattering, in addition to a 3-point interaction involving $\pi_4,\pi_5$ and the dark  gauge field $\mathcal{A}_\mu$ of the gauged~$U(1)_D$, 
\begin{eqnarray} \label{Lkin}
    \mathcal{L}_\mathrm{kin}&=&
    \frac{f_\pi^2}{16}D_\mu\Sigma {D^\mu }\Sigma^\dagger \nn\\&=&\frac{1}{4}\mathrm{Tr}\partial_\mu\pi\partial^\mu\pi
    -   \frac{1}{6f_\pi^2}\mathrm{Tr}\left(\pi^2\partial_\mu\pi\partial^\mu\pi-\pi\partial_\mu\pi\pi\partial^\mu\pi\right)   \nn \\  &&+  \frac{e_D}{4} \mathcal{A}^\mu(\pi_5 \partial_\mu\pi_4 - \pi_4 \partial_\mu\pi_5)+{\cal O}\left(e_D^2,\frac{\pi^6}{f_\pi^4}\right).
\end{eqnarray}
$\mathcal{A}_\mu$ has kinetic mixing with the
$U(1)_Y$ gauge field $B_\mu$~\cite{LEE2015316,Hochberg:2015vrg,Hook:2010tw},
\begin{equation}
    \mathcal{L}_\mathcal{A}=-\frac{1}{4}\mathcal{A}_{\mu\nu}\mathcal{A}^{\mu\nu}-\frac{\sin\chi}{2}B_{\mu\nu}\mathcal{A}^{\mu\nu}+\frac{1}{2}m_A^2 \mathcal{A}_\mu\mathcal{A}^\mu\,,
\end{equation}
with $\mathcal{A}_{\mu\nu}$ and $B_{\mu\nu}$ the respective field strengths. 
When electroweak symmetry breaking occurs,  $\mathcal{A}_\mu$ becomes a mixture of the $Z$-boson and the dark photon~$V$,
\begin{equation}
V^\mu=-\frac{{\rm sin}\zeta}{{\rm cos}\chi}Z^\mu+\frac{{\rm cos}\zeta}{{\rm cos}\chi}\mathcal{A^\mu},
\end{equation}
where 
\begin{equation}
    \tan2\zeta=\frac{m_Z^2\,{\rm sin}\theta_W\,\sin2\chi}{m_A^2-m_Z^2\left({\rm cos}^2\chi-{\rm sin}^2\theta_W {\rm sin}^2\chi\right)}\,,
\end{equation}
and $\theta_W$ the weak mixing angle. 
This generates a $\pi_4f\leftrightarrow\pi_5f$ inelastic scattering process and a $\pi_4\pi_5\rightarrow f \bar{f}$ annihilation process, both mediated by a dark photon $V_\mu$, where $f$ is a SM fermion. For light DM, the leading processes involve abundant electrons; we thus consider this leading contribution of $f=e$ in what follows.

\subsection{SIMPlest Boltzmann Equations}
 The Boltzmann equations governing the pion system include number changing processes:  the $3\rightarrow2$ pion self-annihilation generated by the WZW term and the $\pi_4\pi_5\rightarrow e^+e^-$ co-annihilation that originates from  the kinetic mixing with the SM. The inelastic scattering process ($\pi_4e^\pm\rightarrow\pi_5e^\pm$) generated by the mixing does not alter the total pion number. 
Two-to-two DM self-annihilations will be the last processes to decouple. As a result, all the pions will be in chemical equilibrium, and the heavier pion $\pi_4$ will eventually annihilate to the lighter pions.  
Thus, the number density Boltzmann equation for the pions will be
\begin{eqnarray}\label{BE1pions}
    \dot{n}_{\pi} +3Hn_{\pi}  = &&-\langle \sigma v \rangle_{\mathrm{ann}}^\mathrm{eff}(n_\pi^2 - (n_\pi^\mathrm{eq})^2) \nn\\&&- \langle\sigma v^2\rangle_{3\rightarrow2} \left((n_{\pi})^3 - n_{\pi}^\mathrm{eq}(n_{\pi})^2\right),
\end{eqnarray}
where $n^{(\mathrm{eq})}_\pi=\sum_{i=1}^5 n^{(\mathrm{eq)}}_{\pi_i}$. The $3\rightarrow2$ thermally averaged cross section up to leading order in $\Delta$ is
\begin{equation} \label{32cspions}
    \langle\sigma v^2\rangle_{3\rightarrow2} \simeq\langle\sigma v^2\rangle_{3\rightarrow2}^\mathrm{\Delta=0}\left(1+\frac{14}{125}\Delta\right) 
    \,,
\end{equation}
where $\langle\sigma v^2\rangle_{3\rightarrow2}^\mathrm{\Delta=0}$ is the cross section for the non-broken case calculated in Ref.~\cite{Kuflik:2017iqs}, arising from the WZW term,
\begin{equation} \label{bare32CS}
    \langle\sigma v^2\rangle_{3\rightarrow2}^\mathrm{\Delta=0} = \frac{24\sqrt{5}\tilde{m}_\pi^5}{5\pi^5 f_\pi ^{10}x_\pi^2}.
\end{equation}
The effective co-annihilation process between $\pi_4$ and $\pi_5$  
(up to leading order in $\Delta$) is described by 
\begin{eqnarray} \label{annCS}
    \langle \sigma v \rangle_{\mathrm{ann}}^\mathrm{eff} = \frac{n_{\pi_4}^\mathrm{eq}n_{\pi_5}^\mathrm{eq}}{(n_{\pi}^\mathrm{eq})^2} \langle \sigma v \rangle_{\mathrm{ann}}\simeq e^{-\Delta x_\pi}\frac{\alpha_D\alpha_{\rm EM}\epsilon_\gamma^2\tilde{m}_\pi^2 }{x_\pi m_V^4} 
    \,.
\end{eqnarray}
Here $\langle \sigma v \rangle_{\mathrm{ann}}$ is the $\pi_4\pi_5\rightarrow e^+ e^-$ thermally averaged cross section, $x_\pi=\frac{\tilde{m}_\pi}{T_\pi}$ with $T_\pi$ the temperature of the pions, $m_V$ the mass of the dark photon, $\alpha_{\rm EM}$ ($\alpha_D$) is the (dark) fine structure constant,  
and we define 
\begin{equation}
 \epsilon_\gamma\equiv-{\rm cos}\theta_W {\rm cos}\zeta {\rm tan}\chi
    \end{equation}
    as a dimensionless coefficient related to the mixing angle. 

To evaluate the pion temperature we also need to solve  the energy density Boltzmann equation, 
\beqa\label{eq:BE2pions}
\dot{\rho}_{\pi}+3H(\rho_{\pi}+P_{\pi})&=&n_e^\mathrm{eq}n_{\pi_4}\langle \sigma v\cdot \delta E \rangle_{4\rightarrow5} \no\\ && -  n_e^\mathrm{eq}n_{\pi_5}\langle \sigma v\cdot \delta E \rangle_{5\rightarrow4}\,,
\eeqa
where $\rho_\pi=\sum_{i=1}^5\rho_i, \ P_\pi=\sum_{i=1}^5P_i$ and $\langle \sigma v\cdot \delta E \rangle_{i\rightarrow j}$ is the energy transfer rate generated by the $\pi_4e^\pm\rightarrow\pi_5 e^\pm$ inelastic scattering, with $i,j=4,5$, which can be calculated at leading order in small $\Delta$, large $x=\frac{\tilde{m}_\pi}{T}$ and $x_\pi\sim x$,
\begin{equation} \label{pionET1}
   \langle \sigma v\cdot\delta E\rangle_{4\rightarrow5}\simeq \frac{3 \tilde{m}_\pi m_e^2\alpha_D\alpha_{\rm EM}\epsilon_\gamma^2}{4\pi m_V^4} \frac{ x-x_\pi}{x }\left(1+\frac{\Delta x^3}{4}\right),\
\end{equation}
and 
\begin{equation}\label{pionET2}
       \langle \sigma v\cdot\delta E\rangle_{5\rightarrow4}\simeq e^{-\Delta x} \langle \sigma v\cdot\delta E\rangle_{4\rightarrow5} .
\end{equation}
 
\subsection{SIMPlest Results}
\begin{figure}
    \centering
   \includegraphics[width=0.48\textwidth]{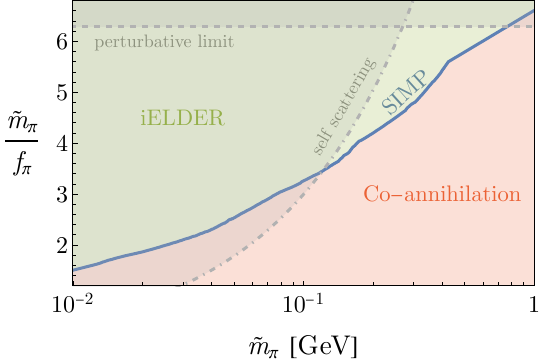}
   \caption{\textbf{Allowed parameter space in the SIMPlest pion model.} 
    The ratio $\tilde m_\pi/f_\pi$
   as a function of the dark pion mass $\tilde m_\pi$, for $m_{V}=3\tilde{m}_\pi, \ \Delta = 0.1$, $\alpha_D=0.1$ and a range of $\epsilon_\gamma$.
    Different colors indicate the mechanism that sets the DM relic abundance across the parameter space: co-annihilation~(shaded red),
    SIMP~(blue curve), and iELDER~(shaded green). 
    The upper shaded gray region indicates the 
    perturbativity limit of $\tilde m_\pi/f_\pi \lesssim 2\pi$,
    while the lower shaded gray region is in tension with the self-scattering constraint of Eq.~\eqref{eq:selfsca}.
   }
    \label{fig:pionParameterSpace}
\end{figure}

Fig.~\ref{fig:PionesPlot} shows our numerical solution to the Boltzmann equations for the pion model, Eqs.~\eqref{BE1pions} and~\eqref{eq:BE2pions}, where we obtain the $\tilde{m}_\pi/f_\pi$ and $\epsilon_\gamma$ relation that produces the observed DM abundance, for various DM masses and mass splittings.  In the left panel, we fix the mass splittings and vary the DM mass, and in the right panel we fix the DM mass and vary the mass splitting. 
We find three distinct regimes depending on the decoupling order of the $3\rightarrow2$ self-annihilations, the inelastic scattering and the $\pi_4\pi_5$ annihilations (see also depiction of case ({\it ii}) in Fig.~\ref{fig:scheme}). 
$\pi_4\pi_5$ annihilations always decouple before the inelastic scattering process. At small values of $\tilde{m}_\pi/f_\pi$, the $3\to2$ self-annihilations shut off before $\pi_4\pi_5$ annihilations, resulting in a co-annihilation phase. As one flows to larger values of $\tilde{m}_\pi/f_\pi$, $3\to2$ self-annihilations decouple after $\pi_4\pi_5$ annihilations but before the ineslatic scattering process shuts off, leading to a SIMP regime. Finally, for even larger values of $\tilde{m}_\pi/f_\pi$, inelastic scattering decouples before $3\to2$ self-annihilations, and an iELDER phase emerges. 
For a fixed value of $\Delta$, as we increase the value of the DM mass we shift the phase space accordingly. In the SIMP regime, higher masses correspond to higher values of $\tilde{m}_\pi/f_\pi$, 
and in the other two regimes of co-annihilation and iELDER, to higher values of $\epsilon_\gamma$. 
At fixed DM mass, a change in the mass splitting results in a very small change of the SIMP phase solution, with a larger impact on the couplings in the iEDLER and co-annihilation phases. The reason is that 
the $3\rightarrow2$ self-annihilation cross section is weakly dependent on $\Delta$ (as in Eq.~\eqref{32cspions}), whereas the annihilation and inelastic scattering cross sections have an exponential dependence on $\Delta$ (as in Eqs.~\eqref{annCS} and ~\eqref{pionET2}), yielding greater sensitivity to $\Delta$.

Although four–pion interactions do not fix the relic abundance, they will constrain the pion mass range.  
Eqs.~\eqref{firstorder}, \eqref{secondorder}, and~\eqref{Lkin}
generate elastic $2\to2$ self-scatterings among all the pion states. 
After freeze-out, the heavy pions ($\pi_{4}$) will efficiently down-scatter into lighter
species, so only the four light pions contribute to the dark matter self-scattering cross section today, $\sigma_{\mathrm{scatter}}$. 
Astrophysical observations such as the Bullet Cluster collision 
and the shapes of galactic halos ~\cite{Rocha:2012jg,Peter:2012jh,Andrade:2020lqq,Shen:2022sbr,Eckert:2022qia,Gilman:2022ida, Tulin:2017ara}
constrain
\begin{equation}
\label{eq:selfsca}
   \frac{\sigma_{\mathrm{scatter}}}{m_{\mathrm{DM}}}\;\lesssim\;1~\mathrm{cm^{2}/g}.
\end{equation}
Following~\cite{Hochberg:2014kqa}, the self-scattering cross section is
\begin{equation}
    \label{eq:sig_self}
    \sigma_\mathrm{scatter}=\frac{35\tilde{m}_\pi^2}{192\pi f_\pi^4}\,,
\end{equation}
and thus Eq.~\eqref{eq:selfsca} translates into an upper limit on the ratio $\tilde m_{\pi}/f_{\pi}$ for any given mass.

In addition, to this astrophysical constraint, we also consider the chiral perturbation theory
requirement of $\tilde m_{\pi}/f_{\pi}\lesssim 2\pi$. The combined results are shown in
Fig.~\ref{fig:pionParameterSpace}. The shaded band at the top, outlined by the dashed gray curve, indicates where perturbativity is lost, 
while the shaded region outlined by the dot-dashed gray curve 
indicates where the self-interaction
constraint of Eq.~\eqref{eq:selfsca} is not met.
The different mechanisms setting the relic abundance in this parameter space---co-annihilation, SIMP,
and iELDER---are indicated by the three colored regions.
We learn that the viable iELDER parameter space in this pion theory is constrained to the mass range
\begin{equation}
130~\mathrm{MeV}\;\lesssim\;\tilde m_{\pi}\;\lesssim\;800~\mathrm{MeV},
\end{equation}
where both self-interaction bounds and perturbativity are satisfied. For larger values of $m_{\pi}/f_{\pi}$, lattice calculations show that the corrections to the self-scattering are large, but are expected to lower the cross-section relative to the leading order result~\cite{Dengler:2024maq}. Due to the large uncertainty in the self-scattering bound and the large lattice corrections,  we consider Eq.~\eqref{eq:selfsca} as a naive approximate constraint on Eq.~\eqref{eq:sig_self}. 
(Note that in light of the uncertainties,  Eq.~\eqref{eq:sig_self} neglects the second order contributions to the self-interaction rate that appear in Eq.~\eqref{secondorder}.)

\begin{figure}
    \centering
   \includegraphics[width=0.48\textwidth]{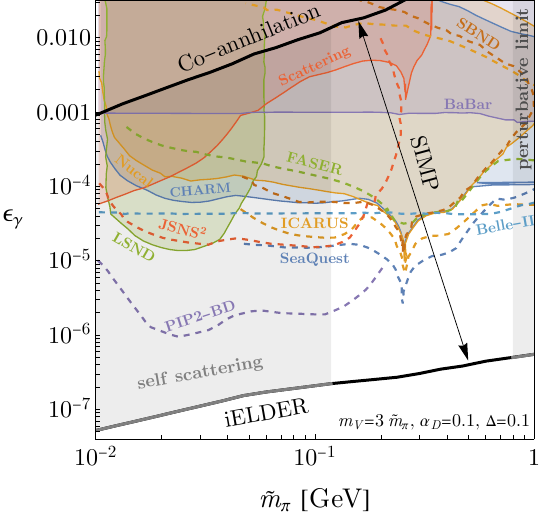}
   \caption{\textbf{SIMPlest pion model phenomenology}. Existing constraints (shaded regions) and future projections (dashed curves) on $\epsilon_\gamma$ and $\tilde{m}_\pi$ for the pion model, with $m_{V}=3\tilde{m}_\pi, \ \Delta = 0.1$ and $\alpha_D=0.1$. 
   Solid black curves indicate the numerical solution to the Boltzmann equations producing the observed DM relic abundance for the co-annihilation and iELDER regimes (upper and lower curves, respectively); the region spanned between these two curves corresponds to the SIMP regime.  
   The parameter space is constrained by data from NuCal and CHARM~\cite{PhysRevLett.126.181801}, LSND~\cite{PhysRevD.98.075020}, BaBar~\cite{PhysRevD.99.015021}, E137, LSND and MiniBooNE~\cite{PhysRevD.96.055007}. Projections are shown for future experiments, including SeaQuest~\cite{PhysRevD.98.035011}, PIP2-BD~\cite{Toups:2022yxs}, short baseline experiments (SBND, MicroBooNE and ICARUS)~\cite{PhysRevD.104.075026}, FASER~\cite{PhysRevD.99.015021}, JNSS$^2$~\cite{PhysRevD.98.075020} and Belle-II~\cite{Duerr_2021}. Gray shaded regions are in tension with self-scattering constraints ({\it left}) and the perturbativity limit ({\it right}).    
   }
    \label{fig:constraints_pions}
\end{figure}
 
 Fig. \ref{fig:constraints_pions} shows the experimental prospects for the SIMPlest pion model presented here in its various phases (see also  Ref.~\cite{Krnjaic:2022ozp}). The parameter space is constrained by data from dark photon decays at beam-dump experiments such as NuCal and CHARM at CERN~\cite{PhysRevLett.126.181801}, LSND~\cite{PhysRevD.98.075020},  missing energy searches at BaBar~\cite{PhysRevD.99.015021}, and down- or up-scattering in E137 at SLAC, LSND and MiniBooNE~\cite{PhysRevD.96.055007}. Future experiments will extend the reach into new regions of parameter space via dark photon decays in beam-dump experiments at Fermilab such as SeaQuest~\cite{PhysRevD.98.035011}, PIP2-BD~\cite{Toups:2022yxs} and the short baseline experiments  (SBND, MicroBooNE and ICARUS)~\cite{PhysRevD.104.075026}, in addition to the JNSS$^2$ beam dump neutrino experiment at J-PARK~\cite{PhysRevD.98.075020}. Other future prospects include decays of the dark photon at particle colliders such as Belle-II at SuperKEKB~\cite{Duerr_2021}, and the FASER detector at the LHC~\cite{PhysRevD.99.015021}.
 Fig.~\ref{fig:constraints_pions} shows the existing constraints (shaded colored regions) and future projections (dashed colored curves) on our parameter space for fixed values of $m_V,\alpha_D$ and $\Delta$, and a range of DM masses and kinetic mixings~$\epsilon_\gamma$.  
 The upper and lower solid black curves indicate  where the numerical solution to the Boltzmann equations produces the observed relic abundance in the co-annihilation and iELDER scenarios, respectively, with the area between the black lines corresponding to the SIMP scenario.  We indicate in shaded gray regions that are in tension with naive limits from self-scattering
 and perturbativity.

While the co-annihilation regime is in general excluded by current experimental measurements for the considered benchmark parameter values, the SIMP regime is partially probed and set to be further covered in upcoming experiments, with the iELDER phase laying just beyond the reach of planned future experiments.

\section{SUMMARY
}\label{sec:out}
 In this work, we introduced the inELastically DEcoupling Relic (iELDER) as a new framework for thermal DM production. 
In this scenario, the DM relic abundance is set by an \textit{inelastic} scattering process---the decoupling of DM--bath scattering---while $3\rightarrow2$ self-annihilations remain active in the dark sector.  Compared to 
co-scattering or elastically decoupling relics, iELDER DM is distinct in its thermal evolution and hierarchy of decoupling processes. iELDER predicts DM in the MeV to GeV mass range along with heavier dark particles not too far in mass, with large $3\to2$ self-annihilations and weak inelastic interactions with bath particles. 

We demonstrated that iELDER DM emerges in scenarios where the DM particle scatters off another particle whose chemical potential remains zero during freeze-out, due to interactions either with the SM bath or with the dark sector. In the first case, we solved the coupled Boltzmann equations for the DM number and energy densities in the presence of inelastic scattering and $3\rightarrow2$ self-annihilations, using a parameterization that is agnostic to the microscopic origin of such interactions. This setup admits a phase diagram partitioned into an iELDER regime and a co-scattering one, with an intermediate transition between the two phases. A toy $Z_3$ theory with effective couplings of the DM to either photons or electrons provided a simplified benchmark model that realized these phases. 
We further presented well-motivated QCD-like theories that realize the second case, utilizing a variation of the dark pion models originally formulated to produce SIMP DM. 
We derived experimental constraints and future projections on this QCD-like theory, leading to interesting phenomenology that can be probed in the near~future.

Incorporating DM inelastic scattering alongside DM self-annihilations offers a rich extension 
to the thermal DM landscape, with phenomenological implications.  Further model-building efforts to construct theories that realize such a scenario would further our understanding of how to constrain or detect iEDLER dark matter.

\textbf{Acknowledgments.} 
We thank Hitoshi Murayama for useful discussions. 
The work of Y.H. is supported by the Israel Science Foundation (grant No. 1818/22), by the Binational Science Foundation (grant No. 2022287) and by an ERC STG grant (``Light-Dark'', grant No. 101040019).  EK is supported by NSF-BSF (grants No. 2022713 and No. 2023711) and by the US-Israeli Binational Science Foundation (grant No.
2020220). B.V. acknowledges support by the Israel Science Foundation (grant No. 1818/22). 
 This project has received funding from the European Research Council (ERC) under the European Union’s Horizon Europe research and innovation programme (grant agreement No. 101040019).  Views and opinions expressed are however those of the author(s) only and do not necessarily reflect those of the European Union. The European Union cannot be held responsible for them.

\bibliography{references}  

\begin{thebibliography}{46}%
\makeatletter
\providecommand \@ifxundefined [1]{%
 \@ifx{#1\undefined}
}%
\providecommand \@ifnum [1]{%
 \ifnum #1\expandafter \@firstoftwo
 \else \expandafter \@secondoftwo
 \fi
}%
\providecommand \@ifx [1]{%
 \ifx #1\expandafter \@firstoftwo
 \else \expandafter \@secondoftwo
 \fi
}%
\providecommand \natexlab [1]{#1}%
\providecommand \enquote  [1]{``#1''}%
\providecommand \bibnamefont  [1]{#1}%
\providecommand \bibfnamefont [1]{#1}%
\providecommand \citenamefont [1]{#1}%
\providecommand \href@noop [0]{\@secondoftwo}%
\providecommand \href [0]{\begingroup \@sanitize@url \@href}%
\providecommand \@href[1]{\@@startlink{#1}\@@href}%
\providecommand \@@href[1]{\endgroup#1\@@endlink}%
\providecommand \@sanitize@url [0]{\catcode `\\12\catcode `\$12\catcode
  `\&12\catcode `\#12\catcode `\^12\catcode `\_12\catcode `\%12\relax}%
\providecommand \@@startlink[1]{}%
\providecommand \@@endlink[0]{}%
\providecommand \url  [0]{\begingroup\@sanitize@url \@url }%
\providecommand \@url [1]{\endgroup\@href {#1}{\urlprefix }}%
\providecommand \urlprefix  [0]{URL }%
\providecommand \Eprint [0]{\href }%
\providecommand \doibase [0]{https://doi.org/}%
\providecommand \selectlanguage [0]{\@gobble}%
\providecommand \bibinfo  [0]{\@secondoftwo}%
\providecommand \bibfield  [0]{\@secondoftwo}%
\providecommand \translation [1]{[#1]}%
\providecommand \BibitemOpen [0]{}%
\providecommand \bibitemStop [0]{}%
\providecommand \bibitemNoStop [0]{.\EOS\space}%
\providecommand \EOS [0]{\spacefactor3000\relax}%
\providecommand \BibitemShut  [1]{\csname bibitem#1\endcsname}%
\let\auto@bib@innerbib\@empty
\bibitem [{\citenamefont {Lee}\ and\ \citenamefont
  {Weinberg}(1977)}]{Lee:1977ua}%
  \BibitemOpen
  \bibfield  {author} {\bibinfo {author} {\bibfnamefont {B.~W.}\ \bibnamefont
  {Lee}}\ and\ \bibinfo {author} {\bibfnamefont {S.}~\bibnamefont {Weinberg}},\
  }\bibfield  {title} {\bibinfo {title} {{Cosmological Lower Bound on Heavy
  Neutrino Masses}},\ }\href {https://doi.org/10.1103/PhysRevLett.39.165}
  {\bibfield  {journal} {\bibinfo  {journal} {Phys. Rev. Lett.}\ }\textbf
  {\bibinfo {volume} {39}},\ \bibinfo {pages} {165} (\bibinfo {year}
  {1977})}\BibitemShut {NoStop}%
\bibitem [{\citenamefont {Griest}\ and\ \citenamefont
  {Seckel}(1991)}]{Griest:1990kh}%
  \BibitemOpen
  \bibfield  {author} {\bibinfo {author} {\bibfnamefont {K.}~\bibnamefont
  {Griest}}\ and\ \bibinfo {author} {\bibfnamefont {D.}~\bibnamefont
  {Seckel}},\ }\bibfield  {title} {\bibinfo {title} {{Three exceptions in the
  calculation of relic abundances}},\ }\href
  {https://doi.org/10.1103/PhysRevD.43.3191} {\bibfield  {journal} {\bibinfo
  {journal} {Phys. Rev. D}\ }\textbf {\bibinfo {volume} {43}},\ \bibinfo
  {pages} {3191} (\bibinfo {year} {1991})}\BibitemShut {NoStop}%
\bibitem [{\citenamefont {Carlson}\ \emph {et~al.}(1992)\citenamefont
  {Carlson}, \citenamefont {Machacek},\ and\ \citenamefont
  {Hall}}]{Carlson:1992fn}%
  \BibitemOpen
  \bibfield  {author} {\bibinfo {author} {\bibfnamefont {E.~D.}\ \bibnamefont
  {Carlson}}, \bibinfo {author} {\bibfnamefont {M.~E.}\ \bibnamefont
  {Machacek}},\ and\ \bibinfo {author} {\bibfnamefont {L.~J.}\ \bibnamefont
  {Hall}},\ }\bibfield  {title} {\bibinfo {title} {{Self-interacting dark
  matter}},\ }\href {https://doi.org/10.1086/171833} {\bibfield  {journal}
  {\bibinfo  {journal} {Astrophys. J.}\ }\textbf {\bibinfo {volume} {398}},\
  \bibinfo {pages} {43} (\bibinfo {year} {1992})}\BibitemShut {NoStop}%
\bibitem [{\citenamefont {Hall}\ \emph {et~al.}(2010)\citenamefont {Hall},
  \citenamefont {Jedamzik}, \citenamefont {March-Russell},\ and\ \citenamefont
  {West}}]{Hall:2009bx}%
  \BibitemOpen
  \bibfield  {author} {\bibinfo {author} {\bibfnamefont {L.~J.}\ \bibnamefont
  {Hall}}, \bibinfo {author} {\bibfnamefont {K.}~\bibnamefont {Jedamzik}},
  \bibinfo {author} {\bibfnamefont {J.}~\bibnamefont {March-Russell}},\ and\
  \bibinfo {author} {\bibfnamefont {S.~M.}\ \bibnamefont {West}},\ }\bibfield
  {title} {\bibinfo {title} {{Freeze-In Production of FIMP Dark Matter}},\
  }\href {https://doi.org/10.1007/JHEP03(2010)080} {\bibfield  {journal}
  {\bibinfo  {journal} {JHEP}\ }\textbf {\bibinfo {volume} {03}},\ \bibinfo
  {pages} {080}},\ \Eprint {https://arxiv.org/abs/0911.1120} {arXiv:0911.1120
  [hep-ph]} \BibitemShut {NoStop}%
\bibitem [{\citenamefont {Hochberg}\ \emph {et~al.}(2014)\citenamefont
  {Hochberg}, \citenamefont {Kuflik}, \citenamefont {Volansky},\ and\
  \citenamefont {Wacker}}]{Hochberg:2014dra}%
  \BibitemOpen
  \bibfield  {author} {\bibinfo {author} {\bibfnamefont {Y.}~\bibnamefont
  {Hochberg}}, \bibinfo {author} {\bibfnamefont {E.}~\bibnamefont {Kuflik}},
  \bibinfo {author} {\bibfnamefont {T.}~\bibnamefont {Volansky}},\ and\
  \bibinfo {author} {\bibfnamefont {J.~G.}\ \bibnamefont {Wacker}},\ }\bibfield
   {title} {\bibinfo {title} {{Mechanism for Thermal Relic Dark Matter of
  Strongly Interacting Massive Particles}},\ }\href
  {https://doi.org/10.1103/PhysRevLett.113.171301} {\bibfield  {journal}
  {\bibinfo  {journal} {Phys. Rev. Lett.}\ }\textbf {\bibinfo {volume} {113}},\
  \bibinfo {pages} {171301} (\bibinfo {year} {2014})},\ \Eprint
  {https://arxiv.org/abs/1402.5143} {arXiv:1402.5143 [hep-ph]} \BibitemShut
  {NoStop}%
\bibitem [{\citenamefont {Hochberg}\ \emph {et~al.}(2015)\citenamefont
  {Hochberg}, \citenamefont {Kuflik}, \citenamefont {Murayama}, \citenamefont
  {Volansky},\ and\ \citenamefont {Wacker}}]{Hochberg:2014kqa}%
  \BibitemOpen
  \bibfield  {author} {\bibinfo {author} {\bibfnamefont {Y.}~\bibnamefont
  {Hochberg}}, \bibinfo {author} {\bibfnamefont {E.}~\bibnamefont {Kuflik}},
  \bibinfo {author} {\bibfnamefont {H.}~\bibnamefont {Murayama}}, \bibinfo
  {author} {\bibfnamefont {T.}~\bibnamefont {Volansky}},\ and\ \bibinfo
  {author} {\bibfnamefont {J.~G.}\ \bibnamefont {Wacker}},\ }\bibfield  {title}
  {\bibinfo {title} {{Model for Thermal Relic Dark Matter of Strongly
  Interacting Massive Particles}},\ }\href
  {https://doi.org/10.1103/PhysRevLett.115.021301} {\bibfield  {journal}
  {\bibinfo  {journal} {Phys. Rev. Lett.}\ }\textbf {\bibinfo {volume} {115}},\
  \bibinfo {pages} {021301} (\bibinfo {year} {2015})},\ \Eprint
  {https://arxiv.org/abs/1411.3727} {arXiv:1411.3727 [hep-ph]} \BibitemShut
  {NoStop}%
\bibitem [{\citenamefont {D'Agnolo}\ and\ \citenamefont
  {Ruderman}(2015)}]{DAgnolo:2015ujb}%
  \BibitemOpen
  \bibfield  {author} {\bibinfo {author} {\bibfnamefont {R.~T.}\ \bibnamefont
  {D'Agnolo}}\ and\ \bibinfo {author} {\bibfnamefont {J.~T.}\ \bibnamefont
  {Ruderman}},\ }\bibfield  {title} {\bibinfo {title} {{Light Dark Matter from
  Forbidden Channels}},\ }\href
  {https://doi.org/10.1103/PhysRevLett.115.061301} {\bibfield  {journal}
  {\bibinfo  {journal} {Phys. Rev. Lett.}\ }\textbf {\bibinfo {volume} {115}},\
  \bibinfo {pages} {061301} (\bibinfo {year} {2015})},\ \Eprint
  {https://arxiv.org/abs/1505.07107} {arXiv:1505.07107 [hep-ph]} \BibitemShut
  {NoStop}%
\bibitem [{\citenamefont {Kuflik}\ \emph {et~al.}(2016)\citenamefont {Kuflik},
  \citenamefont {Perelstein}, \citenamefont {Lorier},\ and\ \citenamefont
  {Tsai}}]{Kuflik:2015isi}%
  \BibitemOpen
  \bibfield  {author} {\bibinfo {author} {\bibfnamefont {E.}~\bibnamefont
  {Kuflik}}, \bibinfo {author} {\bibfnamefont {M.}~\bibnamefont {Perelstein}},
  \bibinfo {author} {\bibfnamefont {N.~R.-L.}\ \bibnamefont {Lorier}},\ and\
  \bibinfo {author} {\bibfnamefont {Y.-D.}\ \bibnamefont {Tsai}},\ }\bibfield
  {title} {\bibinfo {title} {{Elastically Decoupling Dark Matter}},\ }\href
  {https://doi.org/10.1103/PhysRevLett.116.221302} {\bibfield  {journal}
  {\bibinfo  {journal} {Phys. Rev. Lett.}\ }\textbf {\bibinfo {volume} {116}},\
  \bibinfo {pages} {221302} (\bibinfo {year} {2016})},\ \Eprint
  {https://arxiv.org/abs/1512.04545} {arXiv:1512.04545 [hep-ph]} \BibitemShut
  {NoStop}%
\bibitem [{\citenamefont {Kopp}\ \emph {et~al.}(2016)\citenamefont {Kopp},
  \citenamefont {Liu}, \citenamefont {Slatyer}, \citenamefont {Wang},\ and\
  \citenamefont {Xue}}]{Kopp:2016yji}%
  \BibitemOpen
  \bibfield  {author} {\bibinfo {author} {\bibfnamefont {J.}~\bibnamefont
  {Kopp}}, \bibinfo {author} {\bibfnamefont {J.}~\bibnamefont {Liu}}, \bibinfo
  {author} {\bibfnamefont {T.~R.}\ \bibnamefont {Slatyer}}, \bibinfo {author}
  {\bibfnamefont {X.-P.}\ \bibnamefont {Wang}},\ and\ \bibinfo {author}
  {\bibfnamefont {W.}~\bibnamefont {Xue}},\ }\bibfield  {title} {\bibinfo
  {title} {{Impeded Dark Matter}},\ }\href
  {https://doi.org/10.1007/JHEP12(2016)033} {\bibfield  {journal} {\bibinfo
  {journal} {JHEP}\ }\textbf {\bibinfo {volume} {12}},\ \bibinfo {pages}
  {033}},\ \Eprint {https://arxiv.org/abs/1609.02147} {arXiv:1609.02147
  [hep-ph]} \BibitemShut {NoStop}%
\bibitem [{\citenamefont {Soni}\ and\ \citenamefont
  {Zhang}(2016)}]{Soni:2016gzf}%
  \BibitemOpen
  \bibfield  {author} {\bibinfo {author} {\bibfnamefont {A.}~\bibnamefont
  {Soni}}\ and\ \bibinfo {author} {\bibfnamefont {Y.}~\bibnamefont {Zhang}},\
  }\bibfield  {title} {\bibinfo {title} {{Hidden SU(N) Glueball Dark Matter}},\
  }\href {https://doi.org/10.1103/PhysRevD.93.115025} {\bibfield  {journal}
  {\bibinfo  {journal} {Phys. Rev. D}\ }\textbf {\bibinfo {volume} {93}},\
  \bibinfo {pages} {115025} (\bibinfo {year} {2016})},\ \Eprint
  {https://arxiv.org/abs/1602.00714} {arXiv:1602.00714 [hep-ph]} \BibitemShut
  {NoStop}%
\bibitem [{\citenamefont {D'Agnolo}\ \emph {et~al.}(2017)\citenamefont
  {D'Agnolo}, \citenamefont {Pappadopulo},\ and\ \citenamefont
  {Ruderman}}]{DAgnolo:2017dbv}%
  \BibitemOpen
  \bibfield  {author} {\bibinfo {author} {\bibfnamefont {R.~T.}\ \bibnamefont
  {D'Agnolo}}, \bibinfo {author} {\bibfnamefont {D.}~\bibnamefont
  {Pappadopulo}},\ and\ \bibinfo {author} {\bibfnamefont {J.~T.}\ \bibnamefont
  {Ruderman}},\ }\bibfield  {title} {\bibinfo {title} {{Fourth Exception in the
  Calculation of Relic Abundances}},\ }\href
  {https://doi.org/10.1103/PhysRevLett.119.061102} {\bibfield  {journal}
  {\bibinfo  {journal} {Phys. Rev. Lett.}\ }\textbf {\bibinfo {volume} {119}},\
  \bibinfo {pages} {061102} (\bibinfo {year} {2017})},\ \Eprint
  {https://arxiv.org/abs/1705.08450} {arXiv:1705.08450 [hep-ph]} \BibitemShut
  {NoStop}%
\bibitem [{\citenamefont {Kuflik}\ \emph {et~al.}(2017)\citenamefont {Kuflik},
  \citenamefont {Perelstein}, \citenamefont {Lorier},\ and\ \citenamefont
  {Tsai}}]{Kuflik:2017iqs}%
  \BibitemOpen
  \bibfield  {author} {\bibinfo {author} {\bibfnamefont {E.}~\bibnamefont
  {Kuflik}}, \bibinfo {author} {\bibfnamefont {M.}~\bibnamefont {Perelstein}},
  \bibinfo {author} {\bibfnamefont {N.~R.-L.}\ \bibnamefont {Lorier}},\ and\
  \bibinfo {author} {\bibfnamefont {Y.-D.}\ \bibnamefont {Tsai}},\ }\bibfield
  {title} {\bibinfo {title} {{Phenomenology of ELDER Dark Matter}},\ }\href
  {https://doi.org/10.1007/JHEP08(2017)078} {\bibfield  {journal} {\bibinfo
  {journal} {JHEP}\ }\textbf {\bibinfo {volume} {08}},\ \bibinfo {pages}
  {078}},\ \Eprint {https://arxiv.org/abs/1706.05381} {arXiv:1706.05381
  [hep-ph]} \BibitemShut {NoStop}%
\bibitem [{\citenamefont {D'Agnolo}\ \emph {et~al.}(2020)\citenamefont
  {D'Agnolo}, \citenamefont {Pappadopulo}, \citenamefont {Ruderman},\ and\
  \citenamefont {Wang}}]{DAgnolo:2019zkf}%
  \BibitemOpen
  \bibfield  {author} {\bibinfo {author} {\bibfnamefont {R.~T.}\ \bibnamefont
  {D'Agnolo}}, \bibinfo {author} {\bibfnamefont {D.}~\bibnamefont
  {Pappadopulo}}, \bibinfo {author} {\bibfnamefont {J.~T.}\ \bibnamefont
  {Ruderman}},\ and\ \bibinfo {author} {\bibfnamefont {P.-J.}\ \bibnamefont
  {Wang}},\ }\bibfield  {title} {\bibinfo {title} {{Thermal Relic Targets with
  Exponentially Small Couplings}},\ }\href
  {https://doi.org/10.1103/PhysRevLett.124.151801} {\bibfield  {journal}
  {\bibinfo  {journal} {Phys. Rev. Lett.}\ }\textbf {\bibinfo {volume} {124}},\
  \bibinfo {pages} {151801} (\bibinfo {year} {2020})},\ \Eprint
  {https://arxiv.org/abs/1906.09269} {arXiv:1906.09269 [hep-ph]} \BibitemShut
  {NoStop}%
\bibitem [{\citenamefont {Kim}\ and\ \citenamefont
  {Kuflik}(2019)}]{Kim:2019udq}%
  \BibitemOpen
  \bibfield  {author} {\bibinfo {author} {\bibfnamefont {H.}~\bibnamefont
  {Kim}}\ and\ \bibinfo {author} {\bibfnamefont {E.}~\bibnamefont {Kuflik}},\
  }\bibfield  {title} {\bibinfo {title} {{Superheavy Thermal Dark Matter}},\
  }\href {https://doi.org/10.1103/PhysRevLett.123.191801} {\bibfield  {journal}
  {\bibinfo  {journal} {Phys. Rev. Lett.}\ }\textbf {\bibinfo {volume} {123}},\
  \bibinfo {pages} {191801} (\bibinfo {year} {2019})},\ \Eprint
  {https://arxiv.org/abs/1906.00981} {arXiv:1906.00981 [hep-ph]} \BibitemShut
  {NoStop}%
\bibitem [{\citenamefont {Frumkin}\ \emph
  {et~al.}(2023{\natexlab{a}})\citenamefont {Frumkin}, \citenamefont {Kuflik},
  \citenamefont {Lavie},\ and\ \citenamefont {Silverwater}}]{Frumkin:2022ror}%
  \BibitemOpen
  \bibfield  {author} {\bibinfo {author} {\bibfnamefont {R.}~\bibnamefont
  {Frumkin}}, \bibinfo {author} {\bibfnamefont {E.}~\bibnamefont {Kuflik}},
  \bibinfo {author} {\bibfnamefont {I.}~\bibnamefont {Lavie}},\ and\ \bibinfo
  {author} {\bibfnamefont {T.}~\bibnamefont {Silverwater}},\ }\bibfield
  {title} {\bibinfo {title} {{Roadmap to Thermal Dark Matter beyond the Weakly
  Interacting Dark Matter Unitarity Bound}},\ }\href
  {https://doi.org/10.1103/PhysRevLett.130.171001} {\bibfield  {journal}
  {\bibinfo  {journal} {Phys. Rev. Lett.}\ }\textbf {\bibinfo {volume} {130}},\
  \bibinfo {pages} {171001} (\bibinfo {year} {2023}{\natexlab{a}})},\ \Eprint
  {https://arxiv.org/abs/2207.01635} {arXiv:2207.01635 [hep-ph]} \BibitemShut
  {NoStop}%
\bibitem [{\citenamefont {Frumkin}\ \emph
  {et~al.}(2023{\natexlab{b}})\citenamefont {Frumkin}, \citenamefont
  {Hochberg}, \citenamefont {Kuflik},\ and\ \citenamefont
  {Murayama}}]{Frumkin:2021zng}%
  \BibitemOpen
  \bibfield  {author} {\bibinfo {author} {\bibfnamefont {R.}~\bibnamefont
  {Frumkin}}, \bibinfo {author} {\bibfnamefont {Y.}~\bibnamefont {Hochberg}},
  \bibinfo {author} {\bibfnamefont {E.}~\bibnamefont {Kuflik}},\ and\ \bibinfo
  {author} {\bibfnamefont {H.}~\bibnamefont {Murayama}},\ }\bibfield  {title}
  {\bibinfo {title} {{Thermal Dark Matter from Freeze-Out of Inverse Decays}},\
  }\href {https://doi.org/10.1103/PhysRevLett.130.121001} {\bibfield  {journal}
  {\bibinfo  {journal} {Phys. Rev. Lett.}\ }\textbf {\bibinfo {volume} {130}},\
  \bibinfo {pages} {121001} (\bibinfo {year} {2023}{\natexlab{b}})},\ \Eprint
  {https://arxiv.org/abs/2111.14857} {arXiv:2111.14857 [hep-ph]} \BibitemShut
  {NoStop}%
\bibitem [{\citenamefont {Frumkin}\ \emph {et~al.}(2025)\citenamefont
  {Frumkin}, \citenamefont {Hochberg}, \citenamefont {Kuflik},\ and\
  \citenamefont {Murayama}}]{Frumkin:2025dxq}%
  \BibitemOpen
  \bibfield  {author} {\bibinfo {author} {\bibfnamefont {R.}~\bibnamefont
  {Frumkin}}, \bibinfo {author} {\bibfnamefont {Y.}~\bibnamefont {Hochberg}},
  \bibinfo {author} {\bibfnamefont {E.}~\bibnamefont {Kuflik}},\ and\ \bibinfo
  {author} {\bibfnamefont {H.}~\bibnamefont {Murayama}},\ }\bibfield  {title}
  {\bibinfo {title} {{Phases of Dark Matter from Inverse Decays}},\ }\href@noop
  {} {\  (\bibinfo {year} {2025})},\ \Eprint {https://arxiv.org/abs/2504.16981}
  {arXiv:2504.16981 [hep-ph]} \BibitemShut {NoStop}%
\bibitem [{\citenamefont {Battaglieri}\ \emph {et~al.}(2017)\citenamefont
  {Battaglieri} \emph {et~al.}}]{Battaglieri:2017aum}%
  \BibitemOpen
  \bibfield  {author} {\bibinfo {author} {\bibfnamefont {M.}~\bibnamefont
  {Battaglieri}} \emph {et~al.},\ }\bibfield  {title} {\bibinfo {title} {{US
  Cosmic Visions: New Ideas in Dark Matter 2017: Community Report}},\ }in\
  \href@noop {} {\emph {\bibinfo {booktitle} {{U.S. Cosmic Visions: New Ideas
  in Dark Matter}}}}\ (\bibinfo {year} {2017})\ \Eprint
  {https://arxiv.org/abs/1707.04591} {arXiv:1707.04591 [hep-ph]} \BibitemShut
  {NoStop}%
\bibitem [{\citenamefont {Asadi}\ \emph {et~al.}(2022)\citenamefont {Asadi}
  \emph {et~al.}}]{Asadi:2022njl}%
  \BibitemOpen
  \bibfield  {author} {\bibinfo {author} {\bibfnamefont {P.}~\bibnamefont
  {Asadi}} \emph {et~al.},\ }\bibfield  {title} {\bibinfo {title}
  {{Early-Universe Model Building}},\ }\href@noop {} {\  (\bibinfo {year}
  {2022})},\ \Eprint {https://arxiv.org/abs/2203.06680} {arXiv:2203.06680
  [hep-ph]} \BibitemShut {NoStop}%
\bibitem [{\citenamefont {Hochberg}\ \emph {et~al.}(2016)\citenamefont
  {Hochberg}, \citenamefont {Kuflik},\ and\ \citenamefont
  {Murayama}}]{Hochberg:2015vrg}%
  \BibitemOpen
  \bibfield  {author} {\bibinfo {author} {\bibfnamefont {Y.}~\bibnamefont
  {Hochberg}}, \bibinfo {author} {\bibfnamefont {E.}~\bibnamefont {Kuflik}},\
  and\ \bibinfo {author} {\bibfnamefont {H.}~\bibnamefont {Murayama}},\
  }\bibfield  {title} {\bibinfo {title} {Simp spectroscopy},\ }\bibfield
  {journal} {\bibinfo  {journal} {Journal of High Energy Physics}\ }\textbf
  {\bibinfo {volume} {2016}},\ \href {https://doi.org/10.1007/jhep05(2016)090}
  {10.1007/jhep05(2016)090} (\bibinfo {year} {2016})\BibitemShut {NoStop}%
\bibitem [{\citenamefont {Wess}\ and\ \citenamefont
  {Zumino}(1971)}]{Wess:1971yu}%
  \BibitemOpen
  \bibfield  {author} {\bibinfo {author} {\bibfnamefont {J.}~\bibnamefont
  {Wess}}\ and\ \bibinfo {author} {\bibfnamefont {B.}~\bibnamefont {Zumino}},\
  }\bibfield  {title} {\bibinfo {title} {{Consequences of anomalous Ward
  identities}},\ }\href {https://doi.org/10.1016/0370-2693(71)90582-X}
  {\bibfield  {journal} {\bibinfo  {journal} {Phys. Lett. B}\ }\textbf
  {\bibinfo {volume} {37}},\ \bibinfo {pages} {95} (\bibinfo {year}
  {1971})}\BibitemShut {NoStop}%
\bibitem [{\citenamefont {Witten}(1983{\natexlab{a}})}]{Witten:1983tw}%
  \BibitemOpen
  \bibfield  {author} {\bibinfo {author} {\bibfnamefont {E.}~\bibnamefont
  {Witten}},\ }\bibfield  {title} {\bibinfo {title} {{Global Aspects of Current
  Algebra}},\ }\href {https://doi.org/10.1016/0550-3213(83)90063-9} {\bibfield
  {journal} {\bibinfo  {journal} {Nucl. Phys. B}\ }\textbf {\bibinfo {volume}
  {223}},\ \bibinfo {pages} {422} (\bibinfo {year}
  {1983}{\natexlab{a}})}\BibitemShut {NoStop}%
\bibitem [{\citenamefont {Witten}(1983{\natexlab{b}})}]{Witten:1983tx}%
  \BibitemOpen
  \bibfield  {author} {\bibinfo {author} {\bibfnamefont {E.}~\bibnamefont
  {Witten}},\ }\bibfield  {title} {\bibinfo {title} {{Current Algebra, Baryons,
  and Quark Confinement}},\ }\href
  {https://doi.org/10.1016/0550-3213(83)90064-0} {\bibfield  {journal}
  {\bibinfo  {journal} {Nucl. Phys. B}\ }\textbf {\bibinfo {volume} {223}},\
  \bibinfo {pages} {433} (\bibinfo {year} {1983}{\natexlab{b}})}\BibitemShut
  {NoStop}%
\bibitem [{\citenamefont {Peskin}(1980)}]{Peskin:1980gc}%
  \BibitemOpen
  \bibfield  {author} {\bibinfo {author} {\bibfnamefont {M.~E.}\ \bibnamefont
  {Peskin}},\ }\bibfield  {title} {\bibinfo {title} {{The Alignment of the
  Vacuum in Theories of Technicolor}},\ }\href
  {https://doi.org/10.1016/0550-3213(80)90051-6} {\bibfield  {journal}
  {\bibinfo  {journal} {Nucl. Phys. B}\ }\textbf {\bibinfo {volume} {175}},\
  \bibinfo {pages} {197} (\bibinfo {year} {1980})}\BibitemShut {NoStop}%
\bibitem [{\citenamefont {Preskill}(1981)}]{Preskill:1980mz}%
  \BibitemOpen
  \bibfield  {author} {\bibinfo {author} {\bibfnamefont {J.}~\bibnamefont
  {Preskill}},\ }\bibfield  {title} {\bibinfo {title} {{Subgroup Alignment in
  Hypercolor Theories}},\ }\href {https://doi.org/10.1016/0550-3213(81)90265-0}
  {\bibfield  {journal} {\bibinfo  {journal} {Nucl. Phys. B}\ }\textbf
  {\bibinfo {volume} {177}},\ \bibinfo {pages} {21} (\bibinfo {year}
  {1981})}\BibitemShut {NoStop}%
\bibitem [{\citenamefont {Kosower}(1984)}]{Kosower:1984aw}%
  \BibitemOpen
  \bibfield  {author} {\bibinfo {author} {\bibfnamefont {D.~A.}\ \bibnamefont
  {Kosower}},\ }\bibfield  {title} {\bibinfo {title} {{Symmetry-breaking
  patterns in pseudoreal and real gauge theories}},\ }\href
  {https://doi.org/10.1016/0370-2693(84)91806-9} {\bibfield  {journal}
  {\bibinfo  {journal} {Phys. Lett. B}\ }\textbf {\bibinfo {volume} {144}},\
  \bibinfo {pages} {215} (\bibinfo {year} {1984})}\BibitemShut {NoStop}%
\bibitem [{\citenamefont {Kulkarni}\ \emph {et~al.}(2023)\citenamefont
  {Kulkarni}, \citenamefont {Maas}, \citenamefont {Mee}, \citenamefont
  {Nikolic}, \citenamefont {Pradler},\ and\ \citenamefont
  {Zierler}}]{Kulkarni:2022bvh}%
  \BibitemOpen
  \bibfield  {author} {\bibinfo {author} {\bibfnamefont {S.}~\bibnamefont
  {Kulkarni}}, \bibinfo {author} {\bibfnamefont {A.}~\bibnamefont {Maas}},
  \bibinfo {author} {\bibfnamefont {S.}~\bibnamefont {Mee}}, \bibinfo {author}
  {\bibfnamefont {M.}~\bibnamefont {Nikolic}}, \bibinfo {author} {\bibfnamefont
  {J.}~\bibnamefont {Pradler}},\ and\ \bibinfo {author} {\bibfnamefont
  {F.}~\bibnamefont {Zierler}},\ }\bibfield  {title} {\bibinfo {title}
  {{Low-energy effective description of dark $Sp(4)$ theories}},\ }\href
  {https://doi.org/10.21468/SciPostPhys.14.3.044} {\bibfield  {journal}
  {\bibinfo  {journal} {SciPost Phys.}\ }\textbf {\bibinfo {volume} {14}},\
  \bibinfo {pages} {044} (\bibinfo {year} {2023})},\ \Eprint
  {https://arxiv.org/abs/2202.05191} {arXiv:2202.05191 [hep-ph]} \BibitemShut
  {NoStop}%
\bibitem [{\citenamefont {Lee}\ and\ \citenamefont {Seo}(2015)}]{LEE2015316}%
  \BibitemOpen
  \bibfield  {author} {\bibinfo {author} {\bibfnamefont {H.~M.}\ \bibnamefont
  {Lee}}\ and\ \bibinfo {author} {\bibfnamefont {M.-S.}\ \bibnamefont {Seo}},\
  }\bibfield  {title} {\bibinfo {title} {Communication with simp dark mesons
  via z'-portal},\ }\href
  {https://doi.org/https://doi.org/10.1016/j.physletb.2015.07.013} {\bibfield
  {journal} {\bibinfo  {journal} {Physics Letters B}\ }\textbf {\bibinfo
  {volume} {748}},\ \bibinfo {pages} {316} (\bibinfo {year}
  {2015})}\BibitemShut {NoStop}%
\bibitem [{\citenamefont {Hook}\ \emph {et~al.}(2011)\citenamefont {Hook},
  \citenamefont {Izaguirre},\ and\ \citenamefont {Wacker}}]{Hook:2010tw}%
  \BibitemOpen
  \bibfield  {author} {\bibinfo {author} {\bibfnamefont {A.}~\bibnamefont
  {Hook}}, \bibinfo {author} {\bibfnamefont {E.}~\bibnamefont {Izaguirre}},\
  and\ \bibinfo {author} {\bibfnamefont {J.~G.}\ \bibnamefont {Wacker}},\
  }\bibfield  {title} {\bibinfo {title} {{Model Independent Bounds on Kinetic
  Mixing}},\ }\href {https://doi.org/10.1155/2011/859762} {\bibfield  {journal}
  {\bibinfo  {journal} {Adv. High Energy Phys.}\ }\textbf {\bibinfo {volume}
  {2011}},\ \bibinfo {pages} {859762} (\bibinfo {year} {2011})},\ \Eprint
  {https://arxiv.org/abs/1006.0973} {arXiv:1006.0973 [hep-ph]} \BibitemShut
  {NoStop}%
\bibitem [{\citenamefont {Rocha}\ \emph {et~al.}(2013)\citenamefont {Rocha},
  \citenamefont {Peter}, \citenamefont {Bullock}, \citenamefont {Kaplinghat},
  \citenamefont {Garrison-Kimmel}, \citenamefont {Onorbe},\ and\ \citenamefont
  {Moustakas}}]{Rocha:2012jg}%
  \BibitemOpen
  \bibfield  {author} {\bibinfo {author} {\bibfnamefont {M.}~\bibnamefont
  {Rocha}}, \bibinfo {author} {\bibfnamefont {A.~H.~G.}\ \bibnamefont {Peter}},
  \bibinfo {author} {\bibfnamefont {J.~S.}\ \bibnamefont {Bullock}}, \bibinfo
  {author} {\bibfnamefont {M.}~\bibnamefont {Kaplinghat}}, \bibinfo {author}
  {\bibfnamefont {S.}~\bibnamefont {Garrison-Kimmel}}, \bibinfo {author}
  {\bibfnamefont {J.}~\bibnamefont {Onorbe}},\ and\ \bibinfo {author}
  {\bibfnamefont {L.~A.}\ \bibnamefont {Moustakas}},\ }\bibfield  {title}
  {\bibinfo {title} {{Cosmological Simulations with Self-Interacting Dark
  Matter I: Constant Density Cores and Substructure}},\ }\href
  {https://doi.org/10.1093/mnras/sts514} {\bibfield  {journal} {\bibinfo
  {journal} {Mon. Not. Roy. Astron. Soc.}\ }\textbf {\bibinfo {volume} {430}},\
  \bibinfo {pages} {81} (\bibinfo {year} {2013})},\ \Eprint
  {https://arxiv.org/abs/1208.3025} {arXiv:1208.3025 [astro-ph.CO]}
  \BibitemShut {NoStop}%
\bibitem [{\citenamefont {Peter}\ \emph {et~al.}(2013)\citenamefont {Peter},
  \citenamefont {Rocha}, \citenamefont {Bullock},\ and\ \citenamefont
  {Kaplinghat}}]{Peter:2012jh}%
  \BibitemOpen
  \bibfield  {author} {\bibinfo {author} {\bibfnamefont {A.~H.~G.}\
  \bibnamefont {Peter}}, \bibinfo {author} {\bibfnamefont {M.}~\bibnamefont
  {Rocha}}, \bibinfo {author} {\bibfnamefont {J.~S.}\ \bibnamefont {Bullock}},\
  and\ \bibinfo {author} {\bibfnamefont {M.}~\bibnamefont {Kaplinghat}},\
  }\bibfield  {title} {\bibinfo {title} {{Cosmological Simulations with
  Self-Interacting Dark Matter II: Halo Shapes vs. Observations}},\ }\href
  {https://doi.org/10.1093/mnras/sts535} {\bibfield  {journal} {\bibinfo
  {journal} {Mon. Not. Roy. Astron. Soc.}\ }\textbf {\bibinfo {volume} {430}},\
  \bibinfo {pages} {105} (\bibinfo {year} {2013})},\ \Eprint
  {https://arxiv.org/abs/1208.3026} {arXiv:1208.3026 [astro-ph.CO]}
  \BibitemShut {NoStop}%
\bibitem [{\citenamefont {Andrade}\ \emph {et~al.}(2021)\citenamefont
  {Andrade}, \citenamefont {Fuson}, \citenamefont {Gad-Nasr}, \citenamefont
  {Kong}, \citenamefont {Minor}, \citenamefont {Roberts},\ and\ \citenamefont
  {Kaplinghat}}]{Andrade:2020lqq}%
  \BibitemOpen
  \bibfield  {author} {\bibinfo {author} {\bibfnamefont {K.~E.}\ \bibnamefont
  {Andrade}}, \bibinfo {author} {\bibfnamefont {J.}~\bibnamefont {Fuson}},
  \bibinfo {author} {\bibfnamefont {S.}~\bibnamefont {Gad-Nasr}}, \bibinfo
  {author} {\bibfnamefont {D.}~\bibnamefont {Kong}}, \bibinfo {author}
  {\bibfnamefont {Q.}~\bibnamefont {Minor}}, \bibinfo {author} {\bibfnamefont
  {M.~G.}\ \bibnamefont {Roberts}},\ and\ \bibinfo {author} {\bibfnamefont
  {M.}~\bibnamefont {Kaplinghat}},\ }\bibfield  {title} {\bibinfo {title} {{A
  stringent upper limit on dark matter self-interaction cross-section from
  cluster strong lensing}},\ }\href {https://doi.org/10.1093/mnras/stab3241}
  {\bibfield  {journal} {\bibinfo  {journal} {Mon. Not. Roy. Astron. Soc.}\
  }\textbf {\bibinfo {volume} {510}},\ \bibinfo {pages} {54} (\bibinfo {year}
  {2021})},\ \Eprint {https://arxiv.org/abs/2012.06611} {arXiv:2012.06611
  [astro-ph.CO]} \BibitemShut {NoStop}%
\bibitem [{\citenamefont {Shen}\ \emph {et~al.}(2022)\citenamefont {Shen},
  \citenamefont {Brinckmann}, \citenamefont {Rapetti}, \citenamefont
  {Vogelsberger}, \citenamefont {Mantz}, \citenamefont {Zavala},\ and\
  \citenamefont {Allen}}]{Shen:2022sbr}%
  \BibitemOpen
  \bibfield  {author} {\bibinfo {author} {\bibfnamefont {X.}~\bibnamefont
  {Shen}}, \bibinfo {author} {\bibfnamefont {T.}~\bibnamefont {Brinckmann}},
  \bibinfo {author} {\bibfnamefont {D.}~\bibnamefont {Rapetti}}, \bibinfo
  {author} {\bibfnamefont {M.}~\bibnamefont {Vogelsberger}}, \bibinfo {author}
  {\bibfnamefont {A.}~\bibnamefont {Mantz}}, \bibinfo {author} {\bibfnamefont
  {J.}~\bibnamefont {Zavala}},\ and\ \bibinfo {author} {\bibfnamefont {S.~W.}\
  \bibnamefont {Allen}},\ }\bibfield  {title} {\bibinfo {title} {{X-ray
  morphology of cluster-mass haloes in self-interacting dark matter}},\ }\href
  {https://doi.org/10.1093/mnras/stac2376} {\bibfield  {journal} {\bibinfo
  {journal} {Mon. Not. Roy. Astron. Soc.}\ }\textbf {\bibinfo {volume} {516}},\
  \bibinfo {pages} {1302} (\bibinfo {year} {2022})},\ \Eprint
  {https://arxiv.org/abs/2202.00038} {arXiv:2202.00038 [astro-ph.CO]}
  \BibitemShut {NoStop}%
\bibitem [{\citenamefont {Eckert}\ \emph {et~al.}(2022)\citenamefont {Eckert},
  \citenamefont {Ettori}, \citenamefont {Robertson}, \citenamefont {Massey},
  \citenamefont {Pointecouteau}, \citenamefont {Harvey},\ and\ \citenamefont
  {McCarthy}}]{Eckert:2022qia}%
  \BibitemOpen
  \bibfield  {author} {\bibinfo {author} {\bibfnamefont {D.}~\bibnamefont
  {Eckert}}, \bibinfo {author} {\bibfnamefont {S.}~\bibnamefont {Ettori}},
  \bibinfo {author} {\bibfnamefont {A.}~\bibnamefont {Robertson}}, \bibinfo
  {author} {\bibfnamefont {R.}~\bibnamefont {Massey}}, \bibinfo {author}
  {\bibfnamefont {E.}~\bibnamefont {Pointecouteau}}, \bibinfo {author}
  {\bibfnamefont {D.}~\bibnamefont {Harvey}},\ and\ \bibinfo {author}
  {\bibfnamefont {I.~G.}\ \bibnamefont {McCarthy}},\ }\bibfield  {title}
  {\bibinfo {title} {{Constraints on dark matter self-interaction from the
  internal density profiles of X-COP galaxy clusters}},\ }\href
  {https://doi.org/10.1051/0004-6361/202243205} {\bibfield  {journal} {\bibinfo
   {journal} {Astron. Astrophys.}\ }\textbf {\bibinfo {volume} {666}},\
  \bibinfo {pages} {A41} (\bibinfo {year} {2022})},\ \Eprint
  {https://arxiv.org/abs/2205.01123} {arXiv:2205.01123 [astro-ph.CO]}
  \BibitemShut {NoStop}%
\bibitem [{\citenamefont {Gilman}\ \emph {et~al.}(2023)\citenamefont {Gilman},
  \citenamefont {Zhong},\ and\ \citenamefont {Bovy}}]{Gilman:2022ida}%
  \BibitemOpen
  \bibfield  {author} {\bibinfo {author} {\bibfnamefont {D.}~\bibnamefont
  {Gilman}}, \bibinfo {author} {\bibfnamefont {Y.-M.}\ \bibnamefont {Zhong}},\
  and\ \bibinfo {author} {\bibfnamefont {J.}~\bibnamefont {Bovy}},\ }\bibfield
  {title} {\bibinfo {title} {{Constraining resonant dark matter
  self-interactions with strong gravitational lenses}},\ }\href
  {https://doi.org/10.1103/PhysRevD.107.103008} {\bibfield  {journal} {\bibinfo
   {journal} {Phys. Rev. D}\ }\textbf {\bibinfo {volume} {107}},\ \bibinfo
  {pages} {103008} (\bibinfo {year} {2023})},\ \Eprint
  {https://arxiv.org/abs/2207.13111} {arXiv:2207.13111 [astro-ph.CO]}
  \BibitemShut {NoStop}%
\bibitem [{\citenamefont {Tulin}\ and\ \citenamefont
  {Yu}(2018)}]{Tulin:2017ara}%
  \BibitemOpen
  \bibfield  {author} {\bibinfo {author} {\bibfnamefont {S.}~\bibnamefont
  {Tulin}}\ and\ \bibinfo {author} {\bibfnamefont {H.-B.}\ \bibnamefont {Yu}},\
  }\bibfield  {title} {\bibinfo {title} {{Dark Matter Self-interactions and
  Small Scale Structure}},\ }\href
  {https://doi.org/10.1016/j.physrep.2017.11.004} {\bibfield  {journal}
  {\bibinfo  {journal} {Phys. Rept.}\ }\textbf {\bibinfo {volume} {730}},\
  \bibinfo {pages} {1} (\bibinfo {year} {2018})},\ \Eprint
  {https://arxiv.org/abs/1705.02358} {arXiv:1705.02358 [hep-ph]} \BibitemShut
  {NoStop}%
\bibitem [{\citenamefont {Dengler}\ \emph {et~al.}(2024)\citenamefont
  {Dengler}, \citenamefont {Maas},\ and\ \citenamefont
  {Zierler}}]{Dengler:2024maq}%
  \BibitemOpen
  \bibfield  {author} {\bibinfo {author} {\bibfnamefont {Y.}~\bibnamefont
  {Dengler}}, \bibinfo {author} {\bibfnamefont {A.}~\bibnamefont {Maas}},\ and\
  \bibinfo {author} {\bibfnamefont {F.}~\bibnamefont {Zierler}},\ }\bibfield
  {title} {\bibinfo {title} {{Scattering of dark pions in Sp(4) gauge
  theory}},\ }\href {https://doi.org/10.1103/PhysRevD.110.054513} {\bibfield
  {journal} {\bibinfo  {journal} {Phys. Rev. D}\ }\textbf {\bibinfo {volume}
  {110}},\ \bibinfo {pages} {054513} (\bibinfo {year} {2024})},\ \Eprint
  {https://arxiv.org/abs/2405.06506} {arXiv:2405.06506 [hep-lat]} \BibitemShut
  {NoStop}%
\bibitem [{\citenamefont {Tsai}\ \emph {et~al.}(2021)\citenamefont {Tsai},
  \citenamefont {deNiverville},\ and\ \citenamefont
  {Liu}}]{PhysRevLett.126.181801}%
  \BibitemOpen
  \bibfield  {author} {\bibinfo {author} {\bibfnamefont {Y.-D.}\ \bibnamefont
  {Tsai}}, \bibinfo {author} {\bibfnamefont {P.}~\bibnamefont {deNiverville}},\
  and\ \bibinfo {author} {\bibfnamefont {M.~X.}\ \bibnamefont {Liu}},\
  }\bibfield  {title} {\bibinfo {title} {Dark photon and muon
  $g\ensuremath{-}2$ inspired inelastic dark matter models at the high-energy
  intensity frontier},\ }\href {https://doi.org/10.1103/PhysRevLett.126.181801}
  {\bibfield  {journal} {\bibinfo  {journal} {Phys. Rev. Lett.}\ }\textbf
  {\bibinfo {volume} {126}},\ \bibinfo {pages} {181801} (\bibinfo {year}
  {2021})}\BibitemShut {NoStop}%
\bibitem [{\citenamefont {Jordan}\ \emph {et~al.}(2018)\citenamefont {Jordan},
  \citenamefont {Kahn}, \citenamefont {Krnjaic}, \citenamefont {Moschella},\
  and\ \citenamefont {Spitz}}]{PhysRevD.98.075020}%
  \BibitemOpen
  \bibfield  {author} {\bibinfo {author} {\bibfnamefont {J.~R.}\ \bibnamefont
  {Jordan}}, \bibinfo {author} {\bibfnamefont {Y.}~\bibnamefont {Kahn}},
  \bibinfo {author} {\bibfnamefont {G.}~\bibnamefont {Krnjaic}}, \bibinfo
  {author} {\bibfnamefont {M.}~\bibnamefont {Moschella}},\ and\ \bibinfo
  {author} {\bibfnamefont {J.}~\bibnamefont {Spitz}},\ }\bibfield  {title}
  {\bibinfo {title} {Signatures of pseudo-dirac dark matter at high-intensity
  neutrino experiments},\ }\href {https://doi.org/10.1103/PhysRevD.98.075020}
  {\bibfield  {journal} {\bibinfo  {journal} {Phys. Rev. D}\ }\textbf {\bibinfo
  {volume} {98}},\ \bibinfo {pages} {075020} (\bibinfo {year}
  {2018})}\BibitemShut {NoStop}%
\bibitem [{\citenamefont {Berlin}\ and\ \citenamefont
  {Kling}(2019)}]{PhysRevD.99.015021}%
  \BibitemOpen
  \bibfield  {author} {\bibinfo {author} {\bibfnamefont {A.}~\bibnamefont
  {Berlin}}\ and\ \bibinfo {author} {\bibfnamefont {F.}~\bibnamefont {Kling}},\
  }\bibfield  {title} {\bibinfo {title} {Inelastic dark matter at the lhc
  lifetime frontier: Atlas, cms, lhcb, codex-b, faser, and mathusla},\ }\href
  {https://doi.org/10.1103/PhysRevD.99.015021} {\bibfield  {journal} {\bibinfo
  {journal} {Phys. Rev. D}\ }\textbf {\bibinfo {volume} {99}},\ \bibinfo
  {pages} {015021} (\bibinfo {year} {2019})}\BibitemShut {NoStop}%
\bibitem [{\citenamefont {Izaguirre}\ \emph {et~al.}(2017)\citenamefont
  {Izaguirre}, \citenamefont {Kahn}, \citenamefont {Krnjaic},\ and\
  \citenamefont {Moschella}}]{PhysRevD.96.055007}%
  \BibitemOpen
  \bibfield  {author} {\bibinfo {author} {\bibfnamefont {E.}~\bibnamefont
  {Izaguirre}}, \bibinfo {author} {\bibfnamefont {Y.}~\bibnamefont {Kahn}},
  \bibinfo {author} {\bibfnamefont {G.}~\bibnamefont {Krnjaic}},\ and\ \bibinfo
  {author} {\bibfnamefont {M.}~\bibnamefont {Moschella}},\ }\bibfield  {title}
  {\bibinfo {title} {Testing light dark matter coannihilation with fixed-target
  experiments},\ }\href {https://doi.org/10.1103/PhysRevD.96.055007} {\bibfield
   {journal} {\bibinfo  {journal} {Phys. Rev. D}\ }\textbf {\bibinfo {volume}
  {96}},\ \bibinfo {pages} {055007} (\bibinfo {year} {2017})}\BibitemShut
  {NoStop}%
\bibitem [{\citenamefont {Berlin}\ \emph {et~al.}(2018)\citenamefont {Berlin},
  \citenamefont {Gori}, \citenamefont {Schuster},\ and\ \citenamefont
  {Toro}}]{PhysRevD.98.035011}%
  \BibitemOpen
  \bibfield  {author} {\bibinfo {author} {\bibfnamefont {A.}~\bibnamefont
  {Berlin}}, \bibinfo {author} {\bibfnamefont {S.}~\bibnamefont {Gori}},
  \bibinfo {author} {\bibfnamefont {P.}~\bibnamefont {Schuster}},\ and\
  \bibinfo {author} {\bibfnamefont {N.}~\bibnamefont {Toro}},\ }\bibfield
  {title} {\bibinfo {title} {Dark sectors at the fermilab seaquest
  experiment},\ }\href {https://doi.org/10.1103/PhysRevD.98.035011} {\bibfield
  {journal} {\bibinfo  {journal} {Phys. Rev. D}\ }\textbf {\bibinfo {volume}
  {98}},\ \bibinfo {pages} {035011} (\bibinfo {year} {2018})}\BibitemShut
  {NoStop}%
\bibitem [{\citenamefont {Toups}\ \emph {et~al.}(2022)\citenamefont {Toups}
  \emph {et~al.}}]{Toups:2022yxs}%
  \BibitemOpen
  \bibfield  {author} {\bibinfo {author} {\bibfnamefont {M.}~\bibnamefont
  {Toups}} \emph {et~al.},\ }\bibfield  {title} {\bibinfo {title} {{PIP2-BD:
  GeV Proton Beam Dump at Fermilab's PIP-II Linac}},\ }in\ \href@noop {} {\emph
  {\bibinfo {booktitle} {{Snowmass 2021}}}}\ (\bibinfo {year} {2022})\ \Eprint
  {https://arxiv.org/abs/2203.08079} {arXiv:2203.08079 [hep-ex]} \BibitemShut
  {NoStop}%
\bibitem [{\citenamefont {Batell}\ \emph {et~al.}(2021)\citenamefont {Batell},
  \citenamefont {Berger}, \citenamefont {Darm\'e},\ and\ \citenamefont
  {Frugiuele}}]{PhysRevD.104.075026}%
  \BibitemOpen
  \bibfield  {author} {\bibinfo {author} {\bibfnamefont {B.}~\bibnamefont
  {Batell}}, \bibinfo {author} {\bibfnamefont {J.}~\bibnamefont {Berger}},
  \bibinfo {author} {\bibfnamefont {L.}~\bibnamefont {Darm\'e}},\ and\ \bibinfo
  {author} {\bibfnamefont {C.}~\bibnamefont {Frugiuele}},\ }\bibfield  {title}
  {\bibinfo {title} {Inelastic dark matter at the fermilab short baseline
  neutrino program},\ }\href {https://doi.org/10.1103/PhysRevD.104.075026}
  {\bibfield  {journal} {\bibinfo  {journal} {Phys. Rev. D}\ }\textbf {\bibinfo
  {volume} {104}},\ \bibinfo {pages} {075026} (\bibinfo {year}
  {2021})}\BibitemShut {NoStop}%
\bibitem [{\citenamefont {Duerr}\ \emph {et~al.}(2021)\citenamefont {Duerr},
  \citenamefont {Ferber}, \citenamefont {Garcia-Cely}, \citenamefont {Hearty},\
  and\ \citenamefont {Schmidt-Hoberg}}]{Duerr_2021}%
  \BibitemOpen
  \bibfield  {author} {\bibinfo {author} {\bibfnamefont {M.}~\bibnamefont
  {Duerr}}, \bibinfo {author} {\bibfnamefont {T.}~\bibnamefont {Ferber}},
  \bibinfo {author} {\bibfnamefont {C.}~\bibnamefont {Garcia-Cely}}, \bibinfo
  {author} {\bibfnamefont {C.}~\bibnamefont {Hearty}},\ and\ \bibinfo {author}
  {\bibfnamefont {K.}~\bibnamefont {Schmidt-Hoberg}},\ }\bibfield  {title}
  {\bibinfo {title} {Long-lived dark higgs and inelastic dark matter at belle
  ii},\ }\bibfield  {journal} {\bibinfo  {journal} {Journal of High Energy
  Physics}\ }\textbf {\bibinfo {volume} {2021}},\ \href
  {https://doi.org/10.1007/jhep04(2021)146} {10.1007/jhep04(2021)146} (\bibinfo
  {year} {2021})\BibitemShut {NoStop}%
\bibitem [{\citenamefont {Krnjaic}\ \emph {et~al.}(2022)\citenamefont {Krnjaic}
  \emph {et~al.}}]{Krnjaic:2022ozp}%
  \BibitemOpen
  \bibfield  {author} {\bibinfo {author} {\bibfnamefont {G.}~\bibnamefont
  {Krnjaic}} \emph {et~al.},\ }\bibfield  {title} {\bibinfo {title} {{A
  Snowmass Whitepaper: Dark Matter Production at Intensity-Frontier
  Experiments}},\ }\href@noop {} {\  (\bibinfo {year} {2022})},\ \Eprint
  {https://arxiv.org/abs/2207.00597} {arXiv:2207.00597 [hep-ph]} \BibitemShut
  {NoStop}%
\end{thebibliography}%

\onecolumngrid

\appendix 
\section{Energy Transfer Calculation}\label{app:EnergyTransfer} 
In this Appendix, we lay out the full calculation of the iELDER energy transfer that we use for our results. We will begin with calculating the  energy transfer of the forbidden reaction $\langle \sigma v\cdot E \rangle_{\chi\rightarrow\psi}$  depending on both $\chi$'s temperature~$T_2$ and the bath's temperature~$T$, under the assumptions of case ({\it i}). Then we will obtain the forward reaction energy transfer $\langle \sigma v\cdot E \rangle_{\psi\rightarrow\chi}$ via detailed balance, taking $T_2\rightarrow T$. 

To this end, we need to compute the following integral 
\begin{equation}
  n_\phi^\mathrm{eq}n_{\chi} \langle \sigma v\cdot E \rangle_{\chi\rightarrow\psi} = \int d\Pi_{\chi}d\Pi_{\phi_1}d\Pi_{\psi}d\Pi_{\phi_2}E_{\chi}f_{\chi}f_{\phi_1}(2\pi)^4\delta^{(4)}(p_1+p_2-k_1-k_2)\overline{|\mathcal{M}|}_{\chi\rightarrow\psi}^2\,.
\end{equation}
Here $\overline{|\mathcal{M}|}_{\chi\rightarrow\psi}^2$ is the  squared matrix element averaged over initial and final states
corresponding to the $\chi\phi\rightarrow\psi\phi$ process, $d\Pi_{i}=\frac{g_i d^{3}p_{i}}{\left(2\pi\right)^{3}2E_{i}}$, $p_1$ and $p_2$ are the ingoing momentum of $\chi$ and $\phi_1$ respectively, $k_1$ and $k_2$ the outgoing momentum of $\psi$ and $\phi_2$, and we assume the simple form for the phase space distribution $f_i=e^{-\frac{E_i}{T_i}}$. We use
\begin{equation}
    \intop d\Pi_{i}d\Pi{}_{j}\left(2\pi\right)^{4}\delta^{(4)}\left(p_{1}'+p_{2}'-k_{1}'-k_{2}'\right)=g_{i}g_{j}\frac{\beta(s)}{8\pi}\intop_{\cos\theta=-1}^{\cos\theta=1}\frac{d\cos\theta}{2}\frac{d\phi}{2\pi}\,,
\end{equation}
and so we find 
\begin{equation}
n_\phi^\mathrm{eq}n_{\chi} \langle \sigma v\cdot E \rangle_{\chi\rightarrow\psi} =\intop d\Pi_{\chi}d\Pi_{\phi_{1}}E_{\chi}f_{\chi}f_{\phi_{1}}g_{\psi}g_{\phi_2}\frac{\beta\left(s\right)}{8\pi}\Theta\left(s-m_{\psi}^{2}\right)\intop_{\cos\theta=-1}^{\cos\theta=1}\frac{d\cos\theta}{2}\frac{d\phi}{2\pi}{\overline{|\mathcal{M}|}_{\chi\rightarrow\psi}^2} \,,
\end{equation}
where $\Theta$ stands for the Heaviside step function. 
Here $s$ is the Mandelstam variable $s=(p_1+p_2)^2$,  
\begin{equation}
    \beta(s)=\sqrt{1-\frac{2m_{\psi}^{2}}{s}+\frac{m_{\psi}^{4}}{s^{2}}}
    =\left(1-\frac{m_{\psi}^{2}}{s}\right)\,,
\end{equation}
and $(\theta,\phi)$ are the angles in the center of mass frame. The matrix element for $2\rightarrow2$ scattering does not depend on $\phi$, so the integration over it is trivial, 
\begin{equation}
    n_\phi^\mathrm{eq}n_{\chi} \langle \sigma v\cdot E \rangle_{\chi\rightarrow\psi} =\frac{g_{\chi}g_{\phi_1}g_{\psi}g_{\phi_2}}{8\left(2\pi\right)^{5}}\intop\frac{p_{1}^{2}p_{2}^{2}dp_{1}dp_{2}d\cos\theta_{p_{1}p_{2}}}{E_{\chi}E_{\phi_{1}}}E_{\chi}f_{\chi}f_{\phi_{1}}\beta(s)\Theta\left(s-m_{\psi}^{2}\right)\intop_{\cos\theta=-1}^{\cos\theta=1}\frac{d\cos\theta}{2}\overline{|\mathcal{M}|}_{\chi\rightarrow\psi}^2\,,
\end{equation}
 where $\theta_{p_1 p_2}$ is the angle between $p_1$ and $p_2$. We assume a  form  $\overline{|\mathcal{M}|}_{\chi\rightarrow\psi}^2=c_{nl}s^n\cos^l\theta$ where $n,l\in\mathbb{N}$ and $c_{nl}$ are real coefficients.
We can solve the integral over $\cos\theta$, and by changing variables from $\cos\theta_{p_1 p_2}$ to $s$ using the relation
\begin{equation}
    \begin{array}{c}
      s=\left(\sqrt{m_{\chi}^{2}+p_{1}^{2}}+p_2\right)^{2}-\left(\sqrt{p_{1}^{2}+p_{2}^{2}+2p_{1}p_{2}\cos\theta_{p_{1}p_{2}}}\right)^{2}=\\
      \left(\sqrt{m_{\chi}^{2}+p_{1}^{2}}+p_2\right)^{2}-\left(p_{1}^{2}+p_{2}^{2}+2p_{1}p_{2}\cos\theta_{p_{1}p_{2}}\right)\,.
    \end{array}
\end{equation}
Altogether the integral takes a simpler form,
\begin{equation}
     n_\phi^\mathrm{eq}n_{\chi} \langle \sigma v\cdot E \rangle_{\chi\rightarrow\psi} =\frac{g_{\chi}g_{\phi}g_{\psi}g_{\phi}}{16\left(2\pi\right)^{5}}\frac{c_{nl}}{l+1}\intop\frac{p_{1}^{2}p_{2}^{2}dp_{1}dp_{2}}{E_{\chi}E_{\phi_{1}}}\frac{E_{\chi}f_{\chi}f_{\psi_{1}}}{2p_{1}p_{2}}\int_{s_\mathrm{min}}^{s_\mathrm{max}}\left(s^{n}-m_{\psi}^{2}s^{n-1}\right)\Theta\left(s-m_{\psi}^{2}\right)ds\,,
\end{equation}
where the limits of integration are 
\begin{equation}
     \begin{array}{c}
    s_{\min}=\left(\sqrt{m_{\chi}^{2}+p_{1}^{2}}+\sqrt{m_{\psi}^{2}+p_{2}^{2}}\right)^{2}-\left(p_{1}+p_{2}\right)^{2},\\ s_{\max}=\left(\sqrt{m_{\chi}^{2}+p_{1}^{2}}+\sqrt{m_{\psi}^{2}+p_{2}^{2}}\right)^{2}-\left(p_{1}-p_{2}\right)^{2}\,.
     \end{array}
\end{equation}
The integral is
\begin{equation}
    \int_{s_{\min}}^{s_{\max}}f\left(s\right)\Theta\left(s-m_{\psi}^{2}\right)=F\left(s\right)|_{m_{\psi}^{2}}^{s_{\max}}\Theta\left(s_{\max}-m_{\psi}^{2}\right)-F\left(s\right)|^{s_{\min}}_{m_{\psi}^{2}}\Theta\left(s_{\min}-m_{\psi}^{2}\right)\,,
\end{equation}
with $F(s) =\int\left(s^{n}-m_{\psi}^{2}s^{n-1}\right)ds$. In the low temperature limit, we can assume $m_{\psi}^{2}$ is close to $s_{\max}$, so the second term is negligible.  
This was also verified with a numerical calculation of the second term.  We can solve this integral for the simple $n=l=0$ case step by step. First,
\begin{equation}
  F\left(s\right)|_{m_{\psi}^{2}}^{s_{\max}} = -m_{\psi}^{2}+s_{\max} -m_{\psi}^{2}\log\left(\frac{s_{\max}}{m_{\psi}^{2}}\right)\simeq \frac{\left(s_{\max}-m_{\psi}^{2}\right)^{2}}{2m_{\psi}^{2}},
\end{equation}
where we consider only the leading order in $s_{\max}-m_\psi^2$.  (The $n>0$ cases are even simpler because we can proceed without the last Taylor expansion).  Now we solve  the $p_2$ integral, finding eventually a single integral over $p_1$, 
\begin{equation}
     n_\phi^\mathrm{eq}n_{\chi} \langle \sigma v\cdot E \rangle_{\chi\rightarrow\psi} =\frac{g_{\chi}g_{\phi}g_{\psi}g_{\phi}c_{nl}}{16\left(2\pi\right)^{5}}\int_0^\infty\frac{4T^{3}p_{1}\left(m_{2}^{2}+2p_{1}\left(p_{1}+\sqrt{m_{2}^{2}+p_{1}^{2}}\right)\right)}{m_{1}^{2}}e^{-\frac{\left(m_{\psi}^{2}-m_{\chi}^{2}\right)}{2Tm_{\chi}^{2}}\left(\sqrt{m_{2}^{2}+p_{1}^{2}}-p_{1}\right)-\frac{1}{T_{2}}\sqrt{m_{2}^{2}+p_{1}^{2}}}dp_1\,.
\end{equation}
In general, solving this integral analytically is challenging. However, we can get sufficiently close by using a saddle-point approximation on the integrand. As result we finally find 
\begin{equation} \label{fullET} 
    n_\phi^\mathrm{eq}n_{\chi} \langle \sigma v\cdot E \rangle_{\chi\rightarrow\psi}(T,T_2) \propto e^{-\frac{\left(T(m_{\chi}^{2}\left(T-T_{2}\right)+m_{\psi}^{2}T_{2})\right)^{1/2}}{TT_{2}}}.
\end{equation}
This outcome agrees with the numerical calculation of the integral above. For small values of $\Delta$, 
we can further simplify $\langle \sigma v\cdot E \rangle_{\chi\rightarrow\psi} \propto e^{-\Delta x}$ where $x=\frac{m_\chi}{T}$.  We can also see that the result we found above can produce the forward reaction energy transfer $\langle \sigma v\cdot E \rangle_{\psi\rightarrow\chi}$ using detailed balance,
\begin{equation} \label{fullET2}
    \langle \sigma v\cdot E \rangle_{\psi\rightarrow\chi}(T)  = \epsilon^2\left(\frac{m_\psi^4-m_\chi^4}{T m_\psi^4}+\frac{3m_\psi^4+6m_\chi^2m_\psi^2-m_\chi^4}{m_\psi^5}\right)\simeq\ \frac{\epsilon^2}{m_\chi}\left(1+\frac{\Delta x}{2}\right),
\end{equation}
assuming large $x$  and small $\Delta$, with $\epsilon^2= \frac{g_\psi g_\phi c_{nl}}{128\pi}$. A direct calculation of the forward reaction energy transfer under the same assumptions verifies this result. With a similar approach, we can calculate the forward reaction cross-section under the same assumptions to obtain
\begin{equation}
    \langle \sigma v \rangle_{\psi\rightarrow\chi}\simeq \frac{\epsilon^2}{m_\chi^2}\left(1+\frac{\Delta x}{2}\right)\,.
\end{equation}
In solving the coupled Boltzmann equations presented in Eqs.~\eqref{eq:BE1} and~\eqref{eq:BE2} we use the full results as in Eqs.~\eqref{fullET} and~\eqref{fullET2}. However, the leading order in $\Delta x$, presented in Eqs.~\eqref{eq:energyT} and~\eqref{eq:crossS}, is sufficient to understand the roles the parameters play and yields similar results.  
\end{document}